%% file: ms.tex
\documentclass[iop,apj]{emulateapj}
\newcommand{\thirteenco}{$^{13}$CO}
\newcommand{\twelveco}{$^{12}$CO}
\newcommand{\hone}{\ion{H}{1}}
\newcommand{\av}{$A_V$}
\newcommand{\bparallel}{$B_{\parallel}$}
\newcommand{\bturb}{$\langle B_t^2\rangle ^{1/2}/B_0$}
\newcommand{\cf}{\textit{C-F}}

\slugcomment{Accepted in Astrophysical Journal}
\begin{document}

\author{Nicholas L.\ Chapman\altaffilmark{1,2},
Paul F.\ Goldsmith\altaffilmark{1},
Jorge L.\ Pineda\altaffilmark{1},
D.P.\ Clemens\altaffilmark{3},
Di Li\altaffilmark{1},
Marko Kr\v{co}\altaffilmark{4}}

\altaffiltext{1}{Jet Propulsion Laboratory, California Institute of Technology,
4800 Oak Grove Drive, MS 301-429, Pasadena, CA 91109}

\altaffiltext{2}{Center for Interdisciplinary Exploration and Research in
Astrophysics (CIERA), Dept.\ of Physics \& Astronomy, 2145 Sheridan Road,
Evanston, IL 60208; nchapman@u.northwestern.edu}

\altaffiltext{3}{Institute for Astrophysical Research, Boston University,
725 Commonwealth Avenue, Boston, MA 02215}

\altaffiltext{4}{Department of Astronomy, Cornell University, Ithaca, NY 14853}

\title{The Magnetic Field in Taurus Probed by Infrared Polarization}

\begin{abstract}

We present maps of the plane-of-sky magnetic field within two regions of the 
Taurus molecular cloud: one in the dense core L1495/B213 filament, the other in
a diffuse region to the west.  The field is measured from the polarization of
background starlight seen through the cloud.  In total, we measured 287
high-quality near-infrared polarization vectors in these regions.  In  
L1495/B213, the percent polarization increases with column density up to $A_V
\sim 9$ mag, the limits of our data.  The Radiative Torques model for grain
alignment can explain this behavior, but models that invoke turbulence are
inconsistent with the data.  We also combine our data with published optical and
near-infrared polarization measurements in Taurus.  Using this large sample, we
estimate the strength of the plane-of-sky component of the magnetic field in
nine subregions.  This estimation is done with two different techniques that use
the observed dispersion in polarization angles.  Our values range from
$5-82\:\mu$G and tend to be higher in denser regions.  In all subregions, the
critical index of the mass-to-magnetic flux ratio is sub-unity, implying that
Taurus is magnetically supported on large scales ($\sim2$ pc). Within the region
observed, the B213 filament makes a sharp turn to the north and the direction of
the magnetic field also takes a sharp turn, switching from  being perpendicular
to the filament to becoming parallel. This behavior can be understood if we are
observing the rim of a bubble. We argue that it has resulted from a supernova
remnant associated with a recently discovered nearby gamma-ray pulsar.

\end{abstract}


\section{Introduction}

It has been known for decades that observed starlight is polarized
\citep{hall49,hiltner49}.  This polarization is generally understood to be
caused by non-spherical dust grains between Earth and the star that 
preferentially align with their long axes perpendicular to the local magnetic
field direction.  At optical to near-infrared wavelengths, the grains will
preferentially absorb starlight that is linearly polarized parallel to their
long axes.  Thus, it is possible to probe the interstellar magnetic field by
observing the polarized light because at optical and near-infrared wavelengths,
the measured polarization angle will be parallel to the magnetic field
direction.

The importance of magnetic fields to star formation is a hotly debated topic.
Many to most molecular clouds are gravitationally bound, yet the observed star
formation efficiency of clouds is only a few percent \citep{myers86}.  Magnetic
fields are one mechanism that can provide cloud support against gravitational
collapse \citep{mouschovias76}.  An alternative theory is that magnetic fields
are too weak to resist gravity and it is interstellar turbulence that regulates
star formation \citep{maclow04}.

The Taurus molecular cloud is an excellent target for exploring the importance
of magnetic fields in molecular clouds.  As one of the closest low-mass
star-forming clouds at a distance of 140pc
\citep{elias78,kenyon94,wichmann98,loinard05,loinard07,torres07}, Taurus has
been a frequent target for studies \citep[see the extensive review
by][]{kenyon08}.  Furthermore, recent work has revealed a coupling between the
gas and the magnetic field in Taurus.  \citet{goldsmith08} found striations in
their \twelveco{} data that matched the angle of the optical polarization
vectors measured by \citet{moneti84}.  Goldsmith et al.\ offered two possible
mechanisms to explain the striations and their observed alignment with the
magnetic field.  Large-scale flows are seen in the \twelveco{} and \thirteenco{}
channel maps; given the coupled magnetic field and gas, shear in these flows
could explain the striations.  Alternatively, a magnetosonic wave traveling
perpendicular to the field could compress or rarefy the gas into the perceived
striations.  The filamentary region B213 in Taurus also has an apparent
connection between the gas and the magnetic field.  The optical and
near-infrared polarization vectors are oriented perpendicular to the long axis
of the filament \citep{heyer87,goodman92} implying that gravitational collapse
has occurred along the field lines but not across them.

Several studies have measured the large-scale magnetic field in Taurus at
optical wavelengths \citep{moneti84,heyer87,goodman90,whittet92}, but only a
handful of polarization vectors exist in the near-infrared 
\citep{moneti84,tamura87,goodman92}.  Therefore, the magnetic field properties
are poorly known in the high column density regions where stars form. 
Near-infrared observations are needed to penetrate these regions. To address
this deficiency, in this paper we present 287 near-infrared polarization
measurements towards two regions of Taurus.  One region is towards B213 and
L1495 and has not been mapped before in polarization.  Since L1495 is the
highest column density area of Taurus with the highest density of embedded young
stellar objects, measuring the magnetic field in this region is important for
studying the connection between magnetic fields and star formation.  We also
observed a diffuse region near where \citet{goldsmith08} found striations in the
\twelveco{} that matched the angles of the optical polarization vectors.

In \S\,\ref{sec:obs} we discuss the observations, data reduction, and selection
criteria to create the high-quality source catalogs.  We also list references to
previously published optical to near-infrared polarization studies in Taurus
that will augment our new data set.  We use these combined data sets to address
several key magnetic field related questions and issues in
\S\,\ref{sec:results}.  We discuss grain alignment efficiency
(\S\,\ref{sec:align}), magnetic field strength (\S\,\ref{sec:strength}), cloud
stability (\S\,\ref{sec:mu}), turbulence estimates in L1495
(\S\,\ref{sec:turbulence}), and finally the magnetic field morphology in L1495
(\S\,\ref{sec:morphology}).  We summarize our findings in \S\,\ref{sec:summary}.

\section{Observations}\label{sec:obs}

We used the Mimir instrument \citep{clemens07} to observe $H$-band ($1.6\:\mu$m)
polarization of background starlight seen through the Taurus molecular cloud. 
Mimir is installed on the 1.8m Perkins telescope located near Flagstaff, AZ and
operated by Lowell Observatory.  The observations spanned the nights of 2009
Jan.\ 12-16 UT.  We observed two regions within Taurus: one in a low-density
portion of the cloud (hereafter `Diffuse') and the other toward a high-density
region, B213/L1495 (hereafter `Filament').  The Diffuse region is rectangular
covering roughly $4^\mathrm{h}51^\mathrm{m} - 4^\mathrm{h}53^\mathrm{m}$ in
Right Ascension and $25^\circ 26\arcmin - 27^\circ05\arcmin$ in Declination. 
The Filament region is not quite rectangular, but spans approximately
$4^\mathrm{h}16^\mathrm{m}$ to $4^\mathrm{h}19^\mathrm{m}$ in Right Ascension
and $27^\circ11\arcmin - 28^\circ31\arcmin$ in Declination.  Coordinates are
J2000.

\begin{figure*}

\plotone{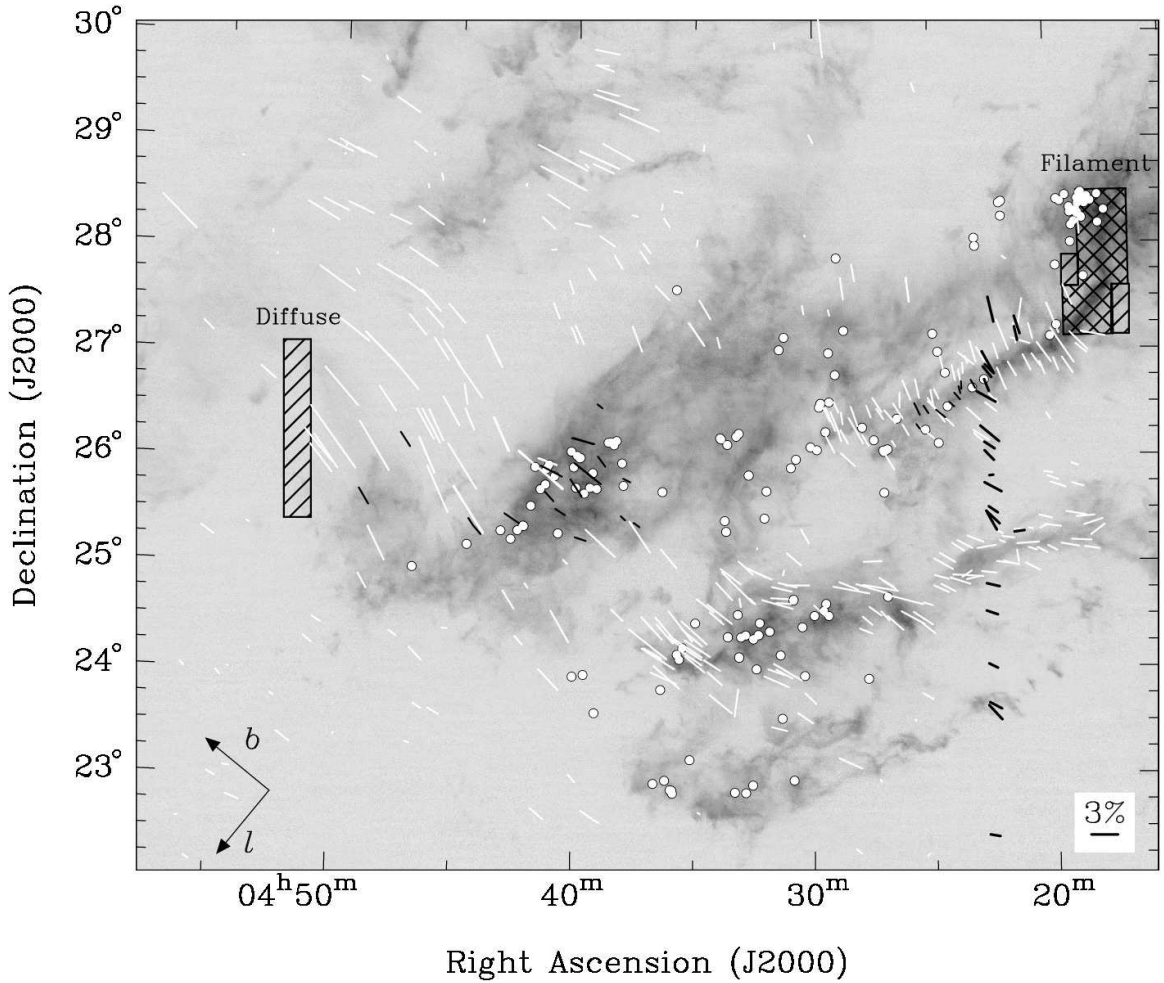}

\caption{\label{fig:map-areas} The two areas mapped are overlaid on the image of
\thirteenco{} emission (integrated over $2-9$ km s$^{-1}$) from
\citet{goldsmith08}.  Areas mapped with short integrations are shown with
slanted lines; those mapped with both short and long integrations are shown
cross-hatched (\S\,\ref{sec:obs}). Also shown are polarization measurements from
previous studies. `Optical' are shown as white lines, `infrared' as black lines,
and `I-band' are shown as thick black lines.  Embedded sources in Taurus are
shown as white circles \citep{luhman06}.  Directions of increasing Galactic
longitude ($l$) and latitude ($b$) are shown in the lower left.}

\end{figure*}

Each region consisted of an overlapping mosaic of $10\arcmin\times10\arcmin$
Mimir fields-of-view. For each FOV, we observed in a six-position hex dither
pattern where each dither position was offset by about $15\arcsec$ from the
center. At each dither position we obtained images at 16 angles of the half-wave
plate (HWP), each separated by $22.5^\circ$.  The integration time per HWP angle
was either 2.5s (short) or 10s (long).  The 16 HWP positions are equivalent to
observing four angles separated by $22.5^\circ$ four times each because the
Stokes parameters vary as $2\theta$ and because polarization ``vectors'' only
have angle and not direction.  Therefore, the effective integration time per
pixel on the sky where all dithers overlap was $4\times6\times2.5$s = 60s
(short) and $4\times6\times10$s = 240s (long).  These times are for each of the
four independent HWP angles.  The short integrations provide polarization
measurements for most sources that were saturated in the long integrations.  The
Diffuse region has low extinction, therefore we observed it only with short
integrations.  For the Filament region, we observed all but five fields with
both short and long integrations.  The two regions observed are indicated in
Figure \ref{fig:map-areas} as hashed (short) and cross-hatched (short and
long).  All previously published optical to near-infrared wavelength
polarization measurements are shown as white and black vectors.  See
\S\,\ref{sec:previous_cats} for details on these catalogs.

We reduced the data using custom IDL programs (Clemens et al.\ in
prep)\footnote{The latest released software can be found
at: http://people.bu.edu/clemens/mimir/software.html }.  Each field, both long
and short integrations, was processed separately to create stellar polarization
lists. We then separately combined the short and long integrations from the
different fields into a single catalog.  For stars that position matched within
$1\arcsec$ a star in adjacent field(s), we computed the weighted average of
their Stokes parameters.  Next we combined our short and long integrations in
the Filament again using a $1\arcsec$ matching radius. For stars that matched we
always used the polarizations obtained from the long integrations, assuming they
were more accurate.  Stars saturated in the long integrations, however, did not
have measured polarizations.  Therefore, when combining the short and long
integrations, these stars have polarizations derived from the short
integrations.

\subsection{High-Quality Vectors\label{sec:quality}}

We applied two selection criteria to the catalogs to produce the high-quality
polarizations used in this paper.  First, we selected only those data with
percentage polarization divided by its uncertainty $p/\sigma_p \geq 3$.  Most of
the sources removed by this criterion are faint and would also be removed by the
second criterion.  The low $p/\sigma_p$ sources have median $p\approx 3\%$ and
median $\sigma_p \approx 2.5\%$.  Secondly, we required sources to
be brighter than 12th mag.\ in the $H$ band for the short integrations and
brighter than 13th mag.\ ($H$ band) for the long integrations.  The second
criterion eliminated the few fainter sources not caught by the first criterion. 
The measured polarizations for fainter sources are unreliable because they are
dominated by errors not accounted for by the first criterion (D.\ Clemens,
private comm.).  These magnitudes are the 2MASS magnitudes for each source. 
Five sources in the Diffuse region did not have 2MASS counterparts and all five
were excluded from the final catalogs.  Nine of the stars in the Filament region
did not have 2MASS counterparts and they were also excluded.

The measured polarization from embedded protostars is not a useful probe of the
magnetic field in molecular clouds.  The radiation from the protostars may drive
dust grain alignment, thus possibly altering the measured polarization.
Furthermore, reflected light from the disk and envelope is highly polarized and
non-uniform.  To avoid this potential bias, we compared our polarization
positions with those of the known protostellar members of Taurus from
\citet{luhman06}.  None of our stars in the Diffuse region position matched
within $1\arcsec$ a known Taurus member.  These 125 sources are listed in Table
\ref{tab:poldiffuse}.  In the Filament, 11 of the 173 polarization measurements
matched within $1\arcsec$ the positions of known Taurus members.  After
excluding these, 162 high-quality polarizations remained and these are listed in
Table \ref{tab:polfilament}.  Because protostars may have some nebulosity 
leading to less certainty in their source positions, we also tried larger 
matching radii, up to $10\arcsec$, and found no difference in the number of 
protostars matched.

\begin{figure*}

\epsscale{0.87}
\plotone{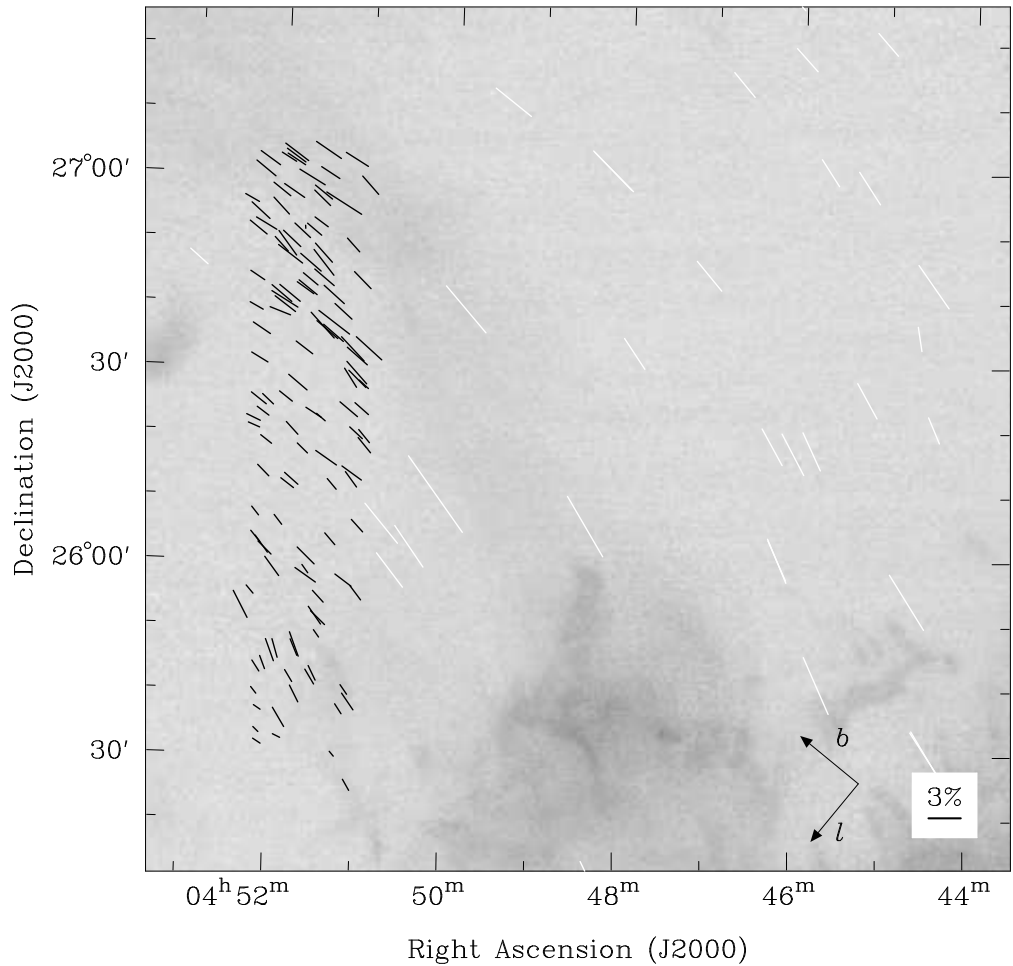}

\caption{\label{fig:map-diffuse} Polarization map of the Diffuse region.
Vectors from Table \ref{tab:poldiffuse} are shown as black lines.  See Fig.\
\ref{fig:map-areas} for a description of the grayscale image and other vectors.
For clarity, infrared vectors from previous studies are not shown.}

\end{figure*}

\begin{figure*}
\epsscale{1}

\plotone{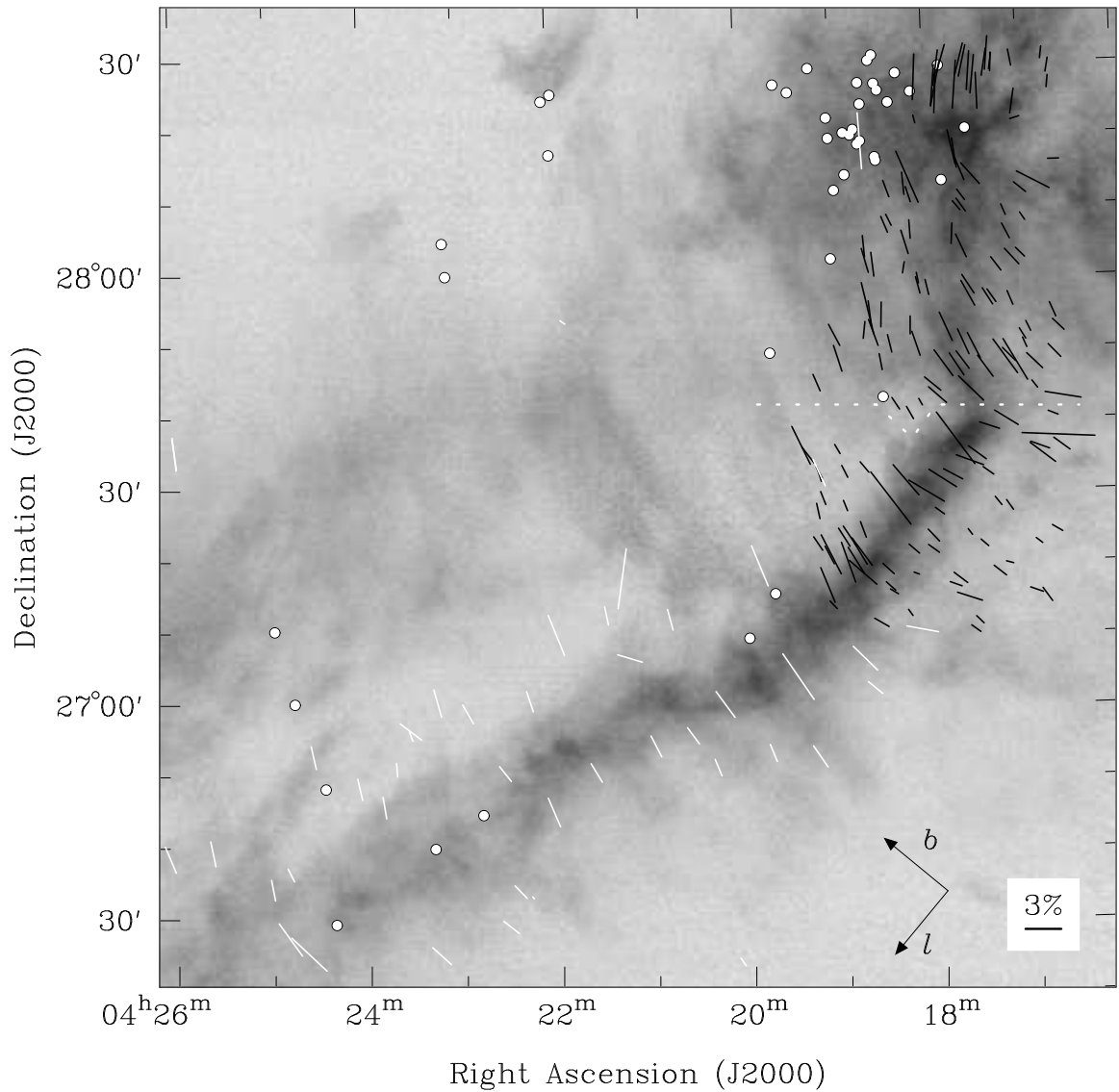}

\caption{\label{fig:map-filament} Polarization map of the Filament region.
Vectors from Table \ref{tab:polfilament} are shown as black lines.  See Fig.\
\ref{fig:map-areas} for a description of the grayscale image, vectors, and white
circles.  For clarity, infrared and I-band vectors from previous studies are not
shown. The dashed white line denotes the division we made between the L1495 and
B213 subregions (Fig.\ \ref{fig:map-subregions}).}

\end{figure*}

Figures \ref{fig:map-diffuse} and \ref{fig:map-filament} show the vectors from
the Diffuse and Filament regions overlaid on the \thirteenco{} image from
\citet{goldsmith08}.  The Mimir near-infrared polarization vectors are in black
and the previously published optical vectors in white.  In Figure
\ref{fig:map-diffuse} the Mimir vectors have polarization angles that are
similar to the optical ones.  In Figure \ref{fig:map-filament}, some of the
Mimir vectors are perpendicular to the B213 filament as are the optical
vectors.  Where the filament abruptly changes direction, the Mimir vectors also
change direction and become oriented north-south.  We include in Figure
\ref{fig:map-filament} a white dotted line separating the B213 filament
(perpendicular vectors) and L1495 (parallel vectors).

\subsection{Previous Polarization Catalogs \label{sec:previous_cats}}

In \S\S\,\ref{sec:strength} and \ref{sec:mu} we will combine previously
published polarization measurements in Taurus along with the Mimir data.  The
published data are identified by wavelength as follows: `optical'
\citep{moneti84,heyer87,goodman90,whittet92}, `infrared'
\citep{moneti84,tamura87,goodman92}, and `I-band' \citep{arce98}.  The I-band
designation is only approximate; the measurements were centered at 7660 \AA,
with a bandpass of 2410 \AA.  For the infrared, we only used $K$-band
polarizations.  Restricting ourselves to $K$ gave us the largest sample since
some stars were only detected at $K$.  We only used sources in these catalogs
with $p/\sigma_p \geq 3$.  Finally, we excluded two vectors from
\citet{goodman92} that are within $1\arcsec$ of known embedded sources
\citep{luhman06}, and two vectors from \citet{arce98} identified by them as
being from stars closer than 150 pc, and thus likely in front of Taurus.

\section{Results}\label{sec:results}

\subsection{Alignment of Dust Grains in Dense Regions\label{sec:align}}

\begin{figure*}

\plotone{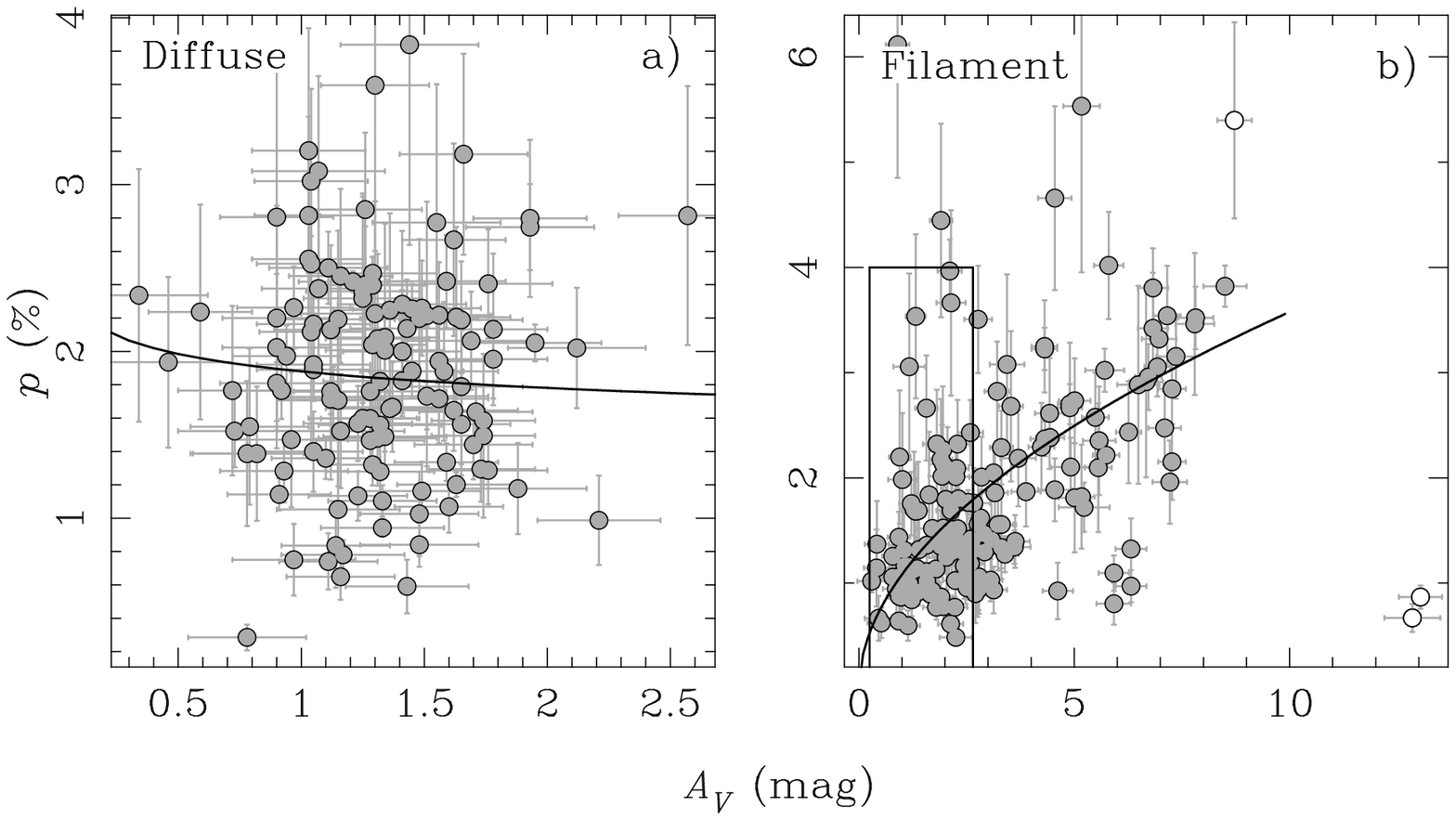}

\caption{\label{fig:av-pol} Percent polarization versus \av{} for: (a) the
Diffuse region and (b) the Filament region.  The solid curves are the
least-squares fit to the data in each region, excluding the open circles in the
Filament region.  The fits are (a) $p = (1.88 \pm 0.09)A_V$$^{-0.08 \pm 0.14}$
and (b) $p = (1.08 \pm 0.06)A_V$$^{0.52 \pm 0.04}$.%
The black rectangle in (b) denotes the range plotted in (a).}

\end{figure*}

To examine how well the dust grains are aligned in different regions of Taurus,
we plot in Figure \ref{fig:av-pol} the percent polarization $p$ versus column
density, \av, for the Mimir sources.  To compute \av, we obtained the 2MASS
$JHK_s$ data for the Taurus cloud and used the \textit{NICER} technique
\citep{lombardi01} to create a $200\arcsec$ resolution extinction map with
$100\arcsec$ pixel spacing \citep{pineda10}.  \textit{NICER} uses the 2MASS
$J-H$ and $H-K_s$ colors simultaneously to estimate the extinction towards each
star.  By using both colors, \textit{NICER} has lower uncertainties than methods
that use $J-H$ or $H-K_s$ alone.  The extinction map is then made by averaging
together the \av{} estimates for stars nearby each pixel.  For each polarization
vector we use the \av{} from the corresponding pixel in the extinction map. 
These values are listed in Tables \ref{tab:poldiffuse} and
\ref{tab:polfilament}.  Estimating \av{} from an extinction map produces more
accurate values of \av{} than just converting the measured color excess to \av{}
for each star because the unknown spectral type for each background star leads
to large uncertainties in \av.

Following \citet{goodman92} and \citet{whittet08}, we fit the data to a power
law $p = a A_V^{\phantom{V}b}$.  Because the errors in \av{} are approximately
all the same, we ignored them to find the best-fit least squares curves for the
two regions, as shown in Figure \ref{fig:av-pol}.  The dynamic range of \av{} in
the Diffuse region is small enough that no correlation between polarization and
\av{} is seen (Pearson correlation coefficient $r=0.05$).  However, in the
Filament region a significant relationship is found, $p=
(1.08\pm0.06)A_V^{\phantom{V}0.52\pm0.04}$, ($r=0.52$).

\begin{figure*}

\plotone{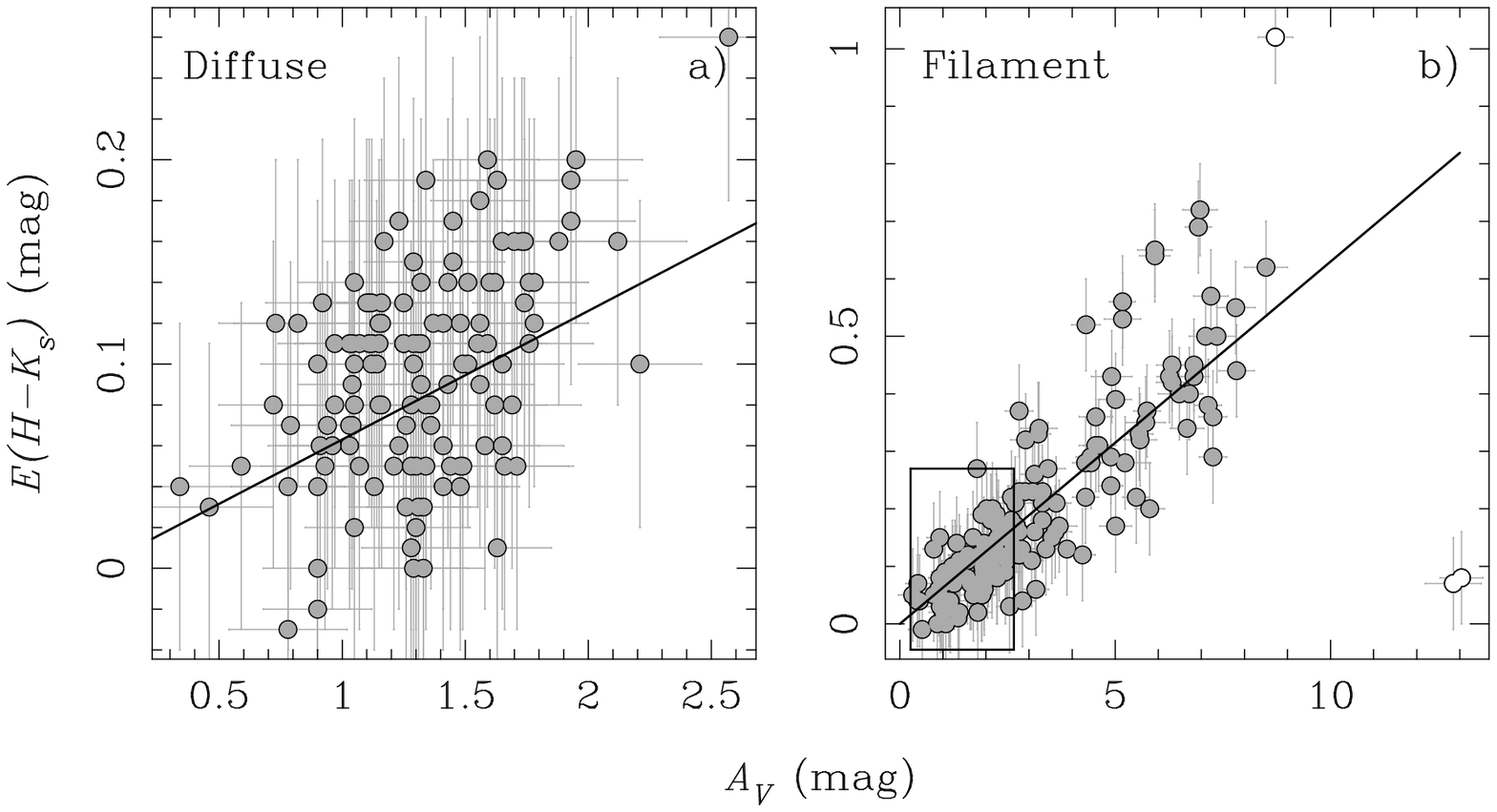}

\caption{\label{fig:comp-avhk} $E(H-K)$ versus \av{} for: (a) the Diffuse region
and (b) the Filament region.  The black lines are the expected \av{} for a given
color excess, $A_V = 15.87E(H-K)$, assuming a standard extinction law
\citep{rieke85}.  The open circles in the Filament are the same sources as in
Fig.\ \ref{fig:av-pol}.  The black rectangle in (b) denotes the range plotted in
(a).}

\end{figure*}

The three open circles in Fig.\ \ref{fig:av-pol} are not used in the fit because
these have \av{} values inconsistent with their measured color excess $E(H-K)$. 
This can be seen in Figure \ref{fig:comp-avhk} where we plot $E(H-K)$ versus
\av{} for the two regions.  To compute $E(H-K)$ we assumed an average intrinsic
color $(H-K) = 0.114 \pm0.074$ as measured by \citet{pineda10}.  The black lines
show the expected extinction for a given $E(H-K)$ assuming a standard dust
extinction law \citep{rieke85}.  In the Diffuse region, all the stars have
$E(H-K)$ within $2\sigma$ of the black line.  However, three sources in the
Filament region are $> 3\sigma$ away from the black line.  The low resolution of
the extinction map has likely caused these sources to be assigned inaccurate
\av{} values. These three sources were thus excluded in the power law fit.

\citet{goodman92,goodman95} and \citet{arce98} measured the near-infrared and
I-band polarization toward several regions in Taurus and Ophiuchus.  They found
that the percentage polarization had a slight dependence on \av, but noted their
data were also consistent with \emph{no} dependence on \av.  They interpreted
these results to argue that while increasing \av{} added more dust grains that
contribute to the column density, these grains did not add polarization.  The
lack of increased polarization with increasing \av{} may be attributed to
changes in dust properties (e.g., formation of icy mantles or grain growth), or
lack of grain alignment in denser regions.  They argued that measurements of the
magnetic field by polarized absorption of background starlight only probe the
magnetic field in a thin surface layer of the cloud.

\citet{whittet08}, however, used $\sim60$ polarizations drawn from many
datasets, including some with extinctions higher than those used by
\citet{goodman92,goodman95} and \citet{arce98}, and found a power law
relationship, $p \propto A_V^{\phantom{V}0.48}$.  They argued that the Radiative
Torques mechanism (RT) for grain alignment
\citep{dolginov76,draine96,draine97,lazarian07} is the best explanation for
their results.  RT assumes there is an anisotropic radiation field impacting the
dust grains.  This radiation will impart a net torque to the grains, assuming
the grains are irregular in shape.  For smooth grains with no helicity the
torque will average out.  The dust grains precess about the magnetic field lines
and one component of the radiative torque is perpendicular to the rotation axis
of the grain and points towards the center of the precession circle, i.e. 
toward the magnetic field line.  On average, this component of the torque acts 
to align the rotation axis of the grain with the magnetic field. 
\citet{whittet08} computed the efficiency for grain alignment by RT with a
simple model.  They assumed a spherical homogeneous cloud with an MRN dust grain
size distribution \citep{mathis77} and with an interstellar radiation field that
matched the average from $0.1-100\:\mu$m \citep{mathis83}.  They then computed
the resultant polarization efficiency versus \av{} and found that RT is capable
of reproducing the observed data up to at least $A_V = 10$ mag.  Beyond this
critical \av{}, $p$ versus \av{} should become flat unless grain growth occurs.

Our power law index in the Filament region, $0.52\pm0.04$, is consistent with
the result obtained by \citeauthor{whittet08}.   Furthermore, our data are
concentrated in a single small region while the data compiled by
\citet{whittet08} span $\sim7^\circ$ in Taurus. The agreement between our
correlation and that found by \citet{whittet08} suggests that RT may be the
primary mechanism for grain alignment in L1495.  Furthermore, because we do not
observe a flattening of the $p$ vs.\ \av{} distribution, it implies the
polarizing dust grains are \emph{not} confined to a thin surface layer of the
cloud, but that grain alignment occurs at least up to \av{} of $\sim9$ mag. This
result is in contrast to \citet{arce98} who found a break point in $p$ vs.\
\av{} beyond which the distribution was flat at $A_V = 1.3\pm0.2$ mag. The
difference in results is possibly explained because the L1495 region has a 
higher column density than the region probed by \citeauthor{arce98} (all their
data have $A_V < 4$ mag).  This is important because the efficiency of RT should
increase with grain size and even more so when the wavelength of the radiation
is comparable to the grain size \citep{cho05,lazarian07}.  Grain growth is a
collisional process, and  so occurs more rapidly in regions with higher number
density.  Furthermore, in higher \av{} regions, shorter wavelengths are
extincted, but infrared photons may still align the large dust grains
\citep{whittet08}.  Lastly, RT relies on anisotropic radiation to work.  Dense
regions will be less fully penetrated by the interstellar radiation field than
more diffuse regions, which may largely neutralize RT in the latter.

Lastly, we should address the possible effect of protostars on our results.
\citeauthor{whittet08} found some evidence for increased polarization efficiency
towards embedded protostars.  It was for this reason that we excluded vectors
from stars that position-matched to known protostars in \S\,\ref{sec:quality}. 
However, the radiation from these protostars could increase the measured
polarization from nearby regions via RT.  We consider a typical low-mass cloud
core radius to be $\sim10,000$ AU \citep[e.g.,][]{benson89}. At the distance of
Taurus 10,000 AU is $\sim70\arcsec$.  We then compared the polarization data in
the Filament region to the embedded star catalog of \citet{luhman06}, using a
$70\arcsec$ matching radius and found only two vectors match (with offset
distances of $28\arcsec$ and $60\arcsec$).  Removing these two sources does not
alter the best-fit relationship between $p$ and \av.  Therefore, the increase in
percent polarization with \av{} is not an artifact caused by embedded
protostellar illumination.

\subsection{ The Magnetic Field Strength\label{sec:strength}}

\begin{figure*}

\plotone{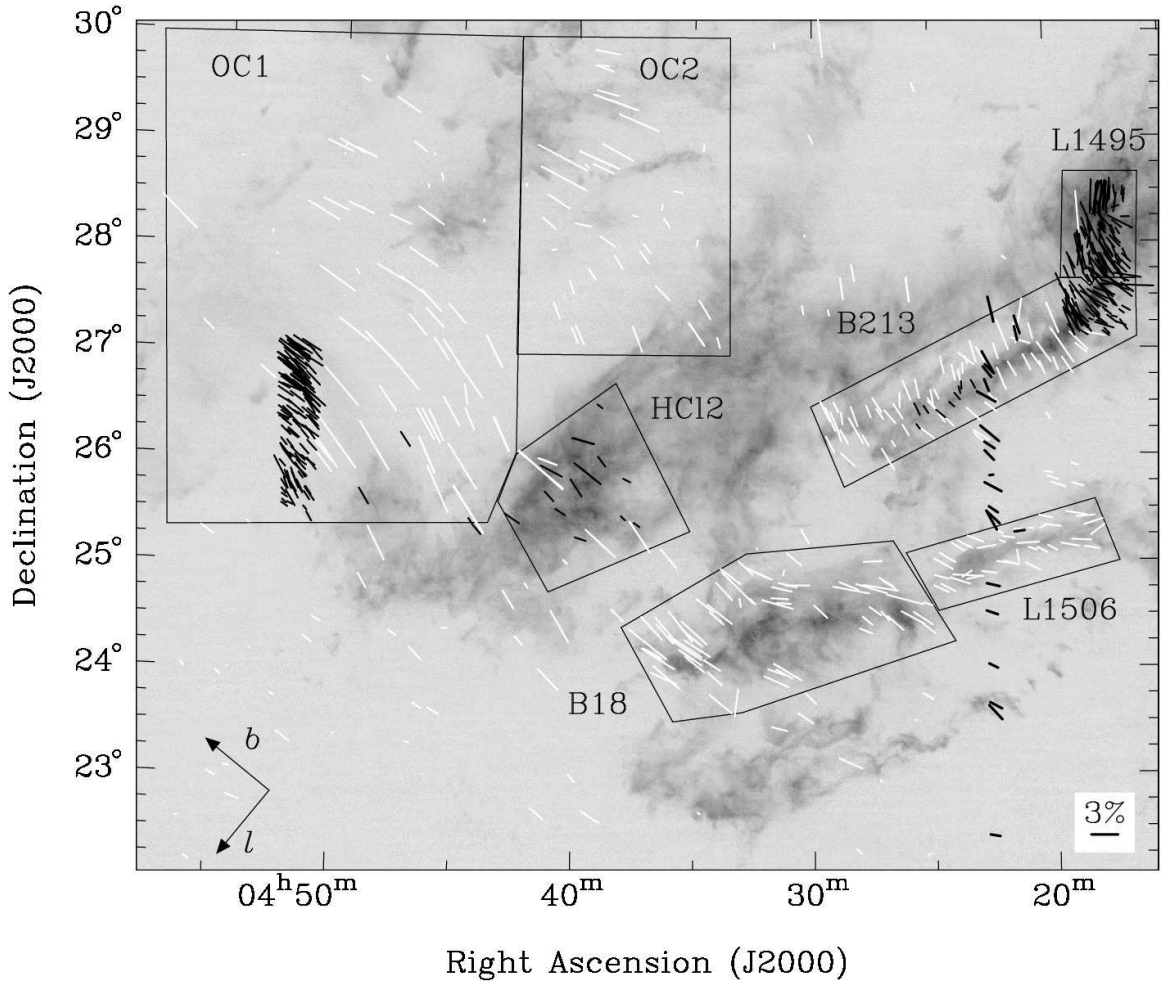}

\caption{\label{fig:map-subregions} Same as Fig.\ \ref{fig:map-areas} with 
Mimir data shown as black vectors.  For clarity, the embedded sources are not
shown.  We divided the map into the subregions labeled here, where HCl2 is
Heiles' Cloud 2, and OC1 and OC2 are two off-cloud regions.}

\end{figure*}

Although the total magnetic field strength, $B$, cannot be measured from our
data, its plane-of-sky component, \bparallel, can be estimated from the
dispersion in angle of the polarization vectors. Under conditions of
flux-freezing, the magnetic field lines should be dragged inward as a region
gravitationally contracts. Therefore, we expect \bparallel{} should increase in
higher density regions. For this reason, we divide the Taurus cloud into the
subregions shown in Figure \ref{fig:map-subregions}. These subregions conform
with known regions in Taurus and the approximate boundaries of the \thirteenco{}
emission.  The exact division between B213 and L1495 can be seen more clearly in
Figure \ref{fig:map-filament}, and corresponds to the Declination where the
filamentary \thirteenco{} emission changes direction  abruptly.  It is not
exactly a straight line because we wanted to make it unambiguous which vectors
were in each subregion.  In addition to the cloud subregions, we chose two
off-cloud subregions denoted OC1 and OC2.  Furthermore, we also use the Diffuse
and Filament Mimir data by themselves to make up a total of nine subregions. 
Note that the Diffuse subregion is just a subset of OC1.

In this section we first present the derivation of the velocity dispersion and
density of the gas in each subregion. These two quantities are needed to compute
\bparallel{}.  Then we will estimate \bparallel{} in each subregion using the
\citet{chandrasekhar53} and \citet{hildebrand09} methods.  The results from this
section are summarized in Tables \ref{tab:regions} and \ref{tab:bfield}.

\input{tabregions.tex}

To estimate the velocity dispersion in each subregion, we first averaged the
\twelveco{} and \thirteenco{} spectrum from each $20\arcsec$ pixel in the
subregion, using the data from \citet{goldsmith08}. We then fit these averaged
\twelveco{} and \thirteenco{} spectra with a gaussian to determine the radial
velocity dispersion, $\sigma(v)$. The line profiles in L1506 have two gaussian
components separated by approximately \mbox{3.5 km s$^{-1}$}.  Therefore, we fit
two gaussians and average their dispersions.  The individual gaussians have
dispersions of 1.05 and 0.87 km s$^{-1}$ for \twelveco{} and 0.47 and 0.52 km
s$^{-1}$ for \thirteenco. The uncertainties in velocity dispersions listed in
Table \ref{tab:regions} are from the fits.  In this paper we use the
\thirteenco{} velocity dispersions to compute \bparallel, as \twelveco{} traces
only a surface layer but \thirteenco{} is sensitive to the whole cloud depth. 
As discussed in \S\,\ref{sec:align}, the polarizing dust does not appear
confined to a thin surface layer.  If we used the \twelveco{} velocity
dispersion instead, all our \bparallel{} estimates would increase by the ratio
of the \twelveco/\thirteenco{} velocity dispersion.

Obtaining the density is more difficult.  Although the extinction map from
\citet{pineda10} can be converted into column density, the number density
remains uncertain since the line-of-sight cloud thicknesses are unknown. 
However, we can estimate the number densities from the CO data. 
\citet{goldsmith08} masked the pixels in the Taurus maps into three regions:
mask0 (neither \twelveco{} nor \thirteenco{} detected in individual pixels),
mask1 (only \twelveco{} detected), and mask2 (both \twelveco{} and \thirteenco{}
detected).  \citet{pineda10} compared the gas and dust in Taurus, including
effects such as depletion, temperature variations, and CO ices.  From their
results, we can use the following estimates for $n(H_2)$: mask0 ($\approx100$
cm$^{-3}$), mask1 (300 cm$^{-3}$), and mask2 ($\geq 10^3$ cm$^{-3}$).  The last
is only a lower limit since $n = 10^3$ cm$^{-3}$ is the critical density of
\thirteenco.  For each polarization vector, we then determine whether it lies in
the mask0, 1, or 2 regions, and use the appropriate number density.  Lastly, we
compute the average $n(H_2)$ for each subregion, using the standard deviation of
the mean to represent the uncertainty.  HCl2 and L1495, both sites with
star-formation, show the highest average densities, close to $10^3$ cm$^{-3}$,
while the off-cloud subregions have lower densities of only a few hundred.

An alternative, more geometric, approach to estimating the density is to assume
that the span on the sky of a subregion is approximately the same as the cloud
depth in that subregion.  We estimate maximum and minimum sizes for each
subregion using the $A_V = 2$ mag.\ contour in the extinction map
\citep{pineda10}. We use the $A_V=2$ mag.\ contour because it outlines the
high-density regions without picking up the low-level diffuse \av.  We then
compute the geometric mean of the maximum and minimum extents to arrive at an
estimate of the cloud depth.  If we assume a standard dust extinction law
\citep{rieke85}, $N(H_2) = 9.4\times 10^{20} A_V$ \citep{bohlin78}.  In all
subregions, the geometric estimates of $n(H_2)$ are less than those obtained
from the masking method.  For approximately square subregions like HCl2 and
L1495, the geometric densities are $\sim1/2$ to 1/3 of the mask-derived
estimates, but for extended regions like B18 and B213, the geometric number
densities are $\sim 1/4$ to $1/3$ of the mask estimates.  Furthermore, the
geometric method cannot be applied to the two off-cloud regions and the Diffuse
region.  For these reasons, in this paper we use density estimates derived from
the CO mask regions.

\subsubsection{ The Chandrasekhar-Fermi Method\label{sec:c-f}}

The method of \citet[hereafter \cf]{chandrasekhar53} estimates the strength of
the plane-of-the-sky component of the magnetic field, \bparallel, using the
measured dispersion in the angles of the polarization vectors.  These authors
assume the dispersion in angles arises from turbulence in the gas and that a
strong magnetic field will resist the turbulence and hence will show a smaller
dispersion in angle.  They derived the expression:

\begin{equation}
B_\parallel = f \frac{\sigma(v)}{\sigma(\theta)}\sqrt{4\pi\rho},
\label{eq:c-f}
\end{equation}

\noindent where $\sigma(\theta)$ is the unweighted dispersion in polarization
angle, $\sigma(v)$ is the line-of-sight velocity dispersion of the gas,  $\rho$
is the gas density, and $f$ is a factor that accounts for averaging of  the
magnetic field direction along the line-of-sight. The last is important  because
along any given line-of-sight there will be many turbulent cells each  with a
different plane-of-sky polarization angle.  The measured polarization  angle at
any point is then the average polarization along the line-of-sight  through the
cloud. This line-of-sight averaging will lead to a smaller measured 
$\sigma(\theta)$, and thus result in an overprediction of \bparallel.

Using Equation \ref{eq:c-f} and the values in Table \ref{tab:regions}, we
estimate \bparallel{} in each subregion.  To compute $\rho$ we convert the
number densities in Table \ref{tab:regions} via the equation $\rho =
n(H_2)m_H\mu_{H_2}$, where $m_H$ is the mass of the hydrogen atom and $\mu_{H_2}
\approx 2.8$ is the mean molecular weight per hydrogen molecule.  Lastly, we set
$f=0.5$.  \citet{ostriker01} found from simulations that this value for $f$ is
a reasonable approximation provided $\sigma(\theta) \lesssim 25^\circ$.  

Uncertainties in \bparallel{} are computed from propagation of errors on
Equation \ref{eq:c-f}.  The propagation of errors is straightforward except for
the uncertainty in $\sigma(\theta)$, which is itself a measured dispersion.  We
go back to the definition of standard deviation and propagate the errors to
obtain the following expression for the uncertainty in $\sigma(\theta)$:

\begin{equation}
\sigma_{\sigma(\theta)} = \frac{1}{(N-1) \sigma(\theta)}\sqrt{\sum \left(
\theta_i - \bar{\theta} \right)^2 \sigma_{\theta_i}^2},
\end{equation}

\noindent where $\theta_i$ and $\sigma_{\theta_i}$ are the angle and
uncertainty for each polarization vector and $\bar{\theta}$ is the average angle
for all the vectors in a subregion.

Table \ref{tab:bfield} lists the values for \bparallel{} in each subregion. The
magnetic field strengths are $10-17\:\mu$G in the low-density off-cloud
subregions, $\sim25\:\mu$G in the B213/Filament/L1495 subregions, and peak at
$42\pm4\:\mu$G in HCl2.  These values are consistent with previous estimates for
magnetic field strengths.  \citet{heyer08} used data from part of OC1 and
estimated $B_\parallel = 14\:\mu$G based on magnetohydrodynamic induced 
velocity anisotropy and separately also from the \cf{} method.   This value is
in excellent  agreement with our values of $17\:\mu$G and $12\:\mu$G for the
diffuse and OC1 regions, respectively.  The line-of-sight component of the
magnetic field, $B_{los}$, has also been measured using OH Zeeman line
splitting.  \citet{troland08} measured OH Zeeman line splitting towards 34
cores, including 11 in Taurus, but only 2 of these 11 with a significance
$\ge3\sigma$. Those two are B217-2 ($B_{los} = 13.5\pm3.7\:\mu$G) and TMC1
($B_{los} = 9.1\pm2.2\:\mu$G).  TMC1 is a core within our HCl2 subregion.  The
much smaller $B_{los}$ measurement compared to \bparallel{} implies that the
magnetic field lies close to the plane-of-the-sky.  However, the relatively
small number of polarization vectors in HCl2 (22) means there may be systematic
errors in the measured value of \bparallel.  

L1506 and OC2 have $\sigma(\theta)$ much larger than $25^\circ$, meaning that
our assumption  of $f=0.5$ may be inaccurate, and thus \bparallel{} is less
certain for these subregions.  L1506 stands out with $\sigma(\theta)=63^\circ$
and the smallest \bparallel. In Figure \ref{fig:map-subregions} the vectors in
the western half of L1506 are approximately parallel to the cloud while those in
the eastern half are more perpendicular, yielding a large $\sigma(\theta)$ and
correspondingly small \bparallel. Future studies should treat these two regions
separately.

\input{tabbfield.tex}

\subsubsection{The Hildebrand et al.\ Method}

The \cf{} method has the advantage of being straightforward to implement, but 
it will yield lower limits for \bparallel{} because it assumes that the 
dispersion in polarization angle is entirely due to turbulence.  Large-scale, 
non-turbulent changes in the magnetic field direction need to be accounted for
because Taurus has large-scale components (see e.g., Fig.\
\ref{fig:map-areas}).  The method of \citet{hildebrand09} accounts for
non-turbulent variations  without assuming a model field.  This method starts by
considering a  two-dimensional map of the magnetic field projected on the
plane-of-the-sky,  where at any position $\mathbf{x}$, the angle of the magnetic
field is  $\Phi(\mathbf{x})$. The difference in angle $\Phi(\mathbf{x}) -
\Phi(\mathbf{x  + \ell})$ is then computed for every pair of vectors.  These
differences in angle are then binned by distance, $\ell$, and the sum over the
$N(\ell)$ pairs of vectors for that bin is computed to arrive at the two-point
correlation (called  the dispersion function):

\begin{equation}
\langle \Delta \Phi^2(\ell)\rangle^{1/2} = \sqrt{\frac{1}{N(\ell)}
\sum_{i=1}^{N(\ell)}[\Phi(\mathbf{x}) - \Phi(\mathbf{x + \ell})]^2}.
\end{equation}

Hildebrand et al.\ assume that $\mathbf{B(x)}$ is composed of a large-scale 
structured field, $\mathbf{B_0(x)}$, and a turbulent component, 
$\mathbf{B_t(x)}$.  Because $\mathbf{B_0(x)}$ is a smoothly varying quantity,
its contribution to the dispersion function should increase linearly with
$\ell$  for small distances.  The turbulent component, $\mathbf{B_b(x)}$, 
should also increase with $\ell$ up to a maximum value when $\ell$ exceeds the
turbulence correlation length $\delta$.  Through Taylor series expansion,
Hildebrand et al.\ separate the contributions from $\mathbf{B_0(x)}$ and
$\mathbf{B_t(x)}$ to the dispersion function.  The authors show that the square
of the dispersion function can be approximated as: 

\begin{equation}
\label{eq:fit}
\langle \Delta \Phi^2(\ell)\rangle_{tot} = b^2 + m^2\ell^2 + \sigma_M^2(\ell),
\end{equation}

\noindent where $\langle \Delta \Phi^2(\ell)\rangle_{tot}$ is the dispersion
function computed from the data.  The quantity $\sigma_M^2(\ell)$ is computed
from the $\Delta \Phi(\ell) = \Phi(\mathbf{x}) - \Phi(\mathbf{x + \ell})$ values
in each bin. Each $\Delta \Phi(\ell)$ has an associated variance obtained from
propagation of errors.  Because the dispersion function is computed by summing
the square of the difference in angles, measurement errors are added in
quadrature and thus will bias the computed $\langle \Delta
\Phi^2(\ell)\rangle_{tot}$ by $\sigma_M^2(\ell)$, where $\sigma_M^2(\ell)$  is
simply the average of the variances on $\Delta \Phi(\ell)$ in a bin.

The quantity $b^2$ is the intercept of a straight line fit to the data (after
subtracting $\sigma_M^2(\ell)$).  As the distance, $\ell$, approaches zero the
contribution to $\langle \Delta \Phi^2(\ell)\rangle$ from $\mathbf{B_0(x)}$
disappears (represented by $m^2$) and only $\mathbf{B_t(x)}$ remains
(represented by $b^2$).  Equation \ref{eq:fit} is valid for displacements $\ell$
larger than the correlation  length for $\mathbf{B_t(x)}$, $\delta$, and for
$\ell$ much smaller than the length scale for variations in $\mathbf{B_0(x)}$,
$d$.

\citeauthor{hildebrand09} then solved their equation for $b^2$ to find the ratio
of the turbulent to the large-scale magnetic field strength:

\begin{equation}
\label{eq:bt}
\frac{\langle B_t^2\rangle ^{1/2}}{B_0} = \frac{b}{\sqrt{2 - b^2}}.
\end{equation}

Lastly, Hildebrand et al.\ assumed that $\sigma(\theta) \simeq \delta B/B_0$,
where $\delta B$ is the variation in magnetic field about the large-scale field
$B_0$.  Therefore, making the inference that $\langle B_t^2\rangle ^{1/2}$ 
corresponds to $\delta B$, Equation \ref{eq:bt} is an expression for
$\sigma(theta)$.

\begin{figure*}

\plotone{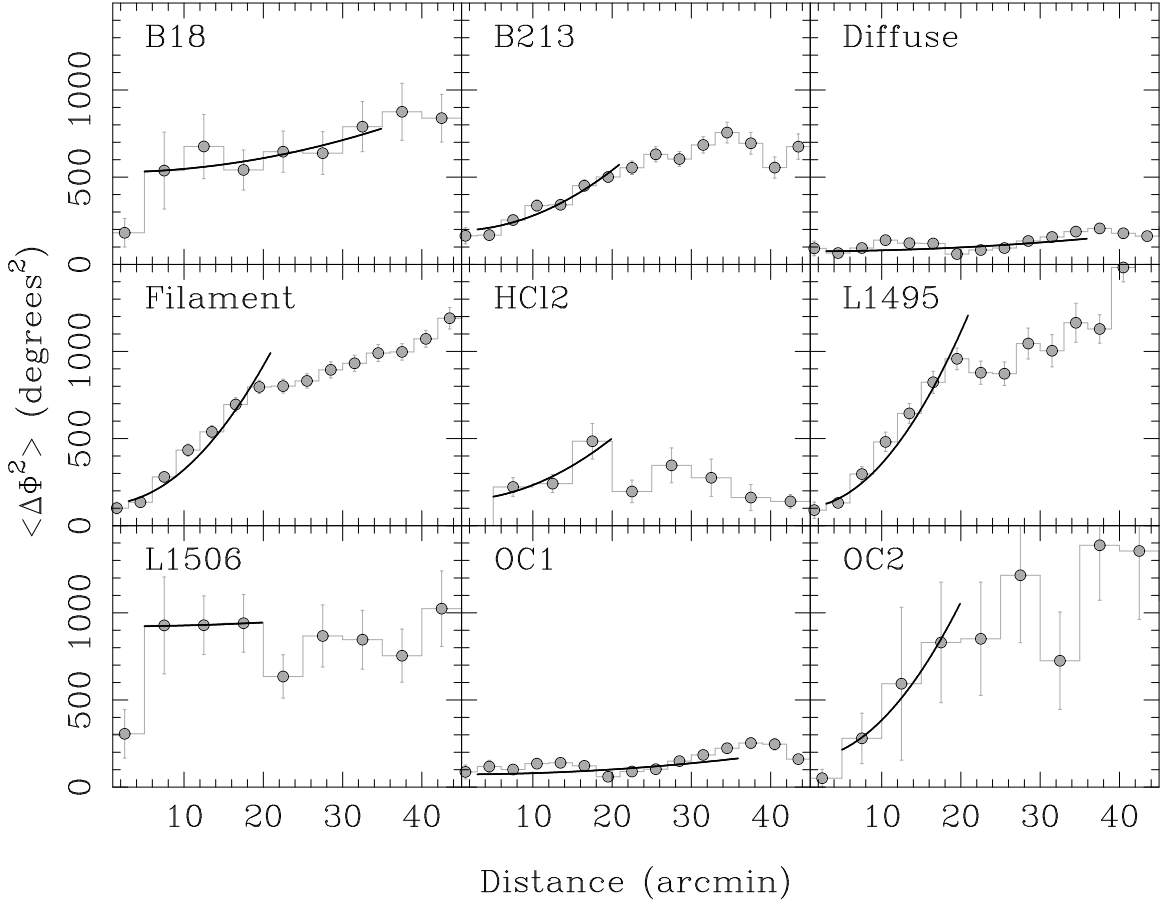}

\caption{\label{fig:bin} Plot of the dispersion function versus distance for the
nine subregions.  The black line shows the best-fit power law to the data
points. Its length denotes the range of distances used in fitting.  We
subtracted $\sigma_M^2(\ell)$ from $\langle \Delta \Phi^2 \rangle$ before
plotting.}

\end{figure*}

We implement the method of \citet{hildebrand09} as follows.  We first  compute
the difference in polarization angle between every set of two points  in a
subregion.  Next, we bin these data into either $3\arcmin$ or  $5\arcmin$ wide
bins and compute the average dispersion function in each bin.   We chose bin
widths of $3\arcmin$ or $5\arcmin$ to have $N(\ell) \ge 10$ for the bins that
are fit with a straight line.  Only one bin has $< 10$ measurements, this is the
third bin in HCl2 with $N(\ell) = 8$.  The five subregions with Mimir data
(B213, Diffuse,  Filament, L1495, and OC1) have many more points than the other
four, allowing us  to use a $3\arcmin$ bin size for these subregions.  Next, we
subtract the average variance, $\sigma_M^2(\ell)$, in each bin.  Lastly, we fit
a straight line to the data versus distance squared.  We always exclude the
first bin because in all subregions this bin has $\sim1/3$ the number of points
compared with the remaining bins.  Furthermore, excluding the first bin ensures
that our distances  will be greater than $\delta$.  \citet{houde09} estimated
$\delta=0.016$ pc towards OMC-1.  At the distance of Taurus 0.016 pc is
$24\arcsec$.  With the exception of three subregions, discussed in the next
paragraph, the maximum distance fit was either $20\arcmin$ or $21\arcmin$,
depending on the bin width.  This maximum distance minimizes the chance that
$\ell > d$. The zero distance intercept of the straight line fit is $b^2$ in
Equation \ref{eq:fit}.  Figure \ref{fig:bin} shows our results from using the
Hildebrand et al.\ method.  The bins are plotted versus distance (not distance
squared) and the uncertainty in each bin is the unweighted standard deviation of
the mean.  The black line shows the best fit and its length represents the range
of distances used in the fit.

Our expectation is that $\langle \Delta \Phi^2(\ell)\rangle$ should increase 
versus offset distance as $\mathbf{B_t(x)}$ and $\mathbf{B_0(x)}$ will cause a
larger dispersion in angles  at larger separation distances \citep[see also
Fig.\ 1 in][]{hildebrand09}.   However, in three subregions the best-fit line
had a negative slope when fitting over the range $5\arcmin-20\arcmin$ (B18) or
$3\arcmin-21\arcmin$ (Diffuse and OC1).  For these three subregions we increased
the range of distances used in fitting until the slope was positive at the
$>1\sigma$ level.  For B18 the  final distances fit are $5\arcmin-35\arcmin$
and for the Diffuse and OC1 subregions the range is $3\arcmin-36\arcmin$. 
These are the ranges shown in Figure \ref{fig:bin}, and the $b^2$ values in
Table \ref{tab:bfield} are  derived from them.  By using the larger distances in
fitting we run a greater risk of $\ell$ exceeding $d$.  The approach we
use may be justified in the Diffuse and OC1 subregions, where from Figure 
\ref{fig:map-areas} it appears that $d$ is greater than $1^\circ$, but it is
unclear if the same is true for B18. However, the important quantity for our
purposes is the intercept, $b^2$, which  is less sensitive to the maximum fit
distance used, giving us confidence in our best-fit values.  In B18, the
best-fit $b^2$ varies by less than the error for maximum fit distances
$25\arcmin-45\arcmin$, and in the Diffuse and OC1 subregions, maximum distances
from $30\arcmin$ to $60\arcmin$ produce best-fit values of $b^2$ that vary from
each other by less than the error listed in Table \ref{tab:bfield}.

With our $b^2$ values, we use Equation \ref{eq:bt} to obtain $\sigma(\theta)$
and  finally Equation \ref{eq:c-f} to estimate \bparallel.  Our results are
listed  in Table \ref{tab:bfield}.  The \citeauthor{hildebrand09} method
produces  estimates of \bparallel{} that are $\sim1.5-4\times$ larger than those
from  \cf.  This increase is expected as the former removes large-scale
$B$-field effects.  In general, denser regions tend to have higher values for
\bparallel.  This behavior is not surprising since $B_\parallel \propto
\sqrt{\rho}$ from Equation \ref{eq:c-f}, but may also indicate that frozen
magnetic field lines are being dragged inward by gravity.

The relatively limited number of vectors in regions without Mimir data may
introduce biases in our estimates of \bparallel{}.  For example, L1506 has an
extremely high value for $b^2$ that is determined  to high precision, leading to
a well-determined estimate for \bparallel{} that is much lower than what is
found even in low-density regions ($12\pm1\:\mu$G).  Another possible source of
systematic error is the correction factor $f$.  Because \citet{hildebrand09}
accounts for non-turbulent variations in magnetic field direction, it is
possible that $f\approx0.5$ even in regions where $\sigma(\theta) \gtrsim 
25^\circ$.  It is also possible that a different correction factor is needed for
the Hildebrand et al.\ technique compared to \cf. More work is needed to
understand how $f$ depends on physical quantities.

\subsection{ Cloud Stability\label{sec:mu}}

Is the magnetic field strength in Taurus sufficiently strong to be important in
regulating star formation?  \citet{nakano78} derived the stability criterion for
an isothermal gaseous layer threaded by a perpendicular magnetic field to be
$(M/\Phi)_\mathrm{crit} = 1/\sqrt{4 \pi^2 G}$.  For a mass-to-magnetic flux
ratio exceeding this value, the region will collapse owing to gravity.  The
stability criterion can be rewritten in terms of the  dimensionless magnetic
critical index:

\begin{equation}
\label{eq:mu1}
\mu = \frac{(M/\Phi)}{(M/\Phi)_\mathrm{crit}} = 7.6 N_\parallel(H_2)/B_{tot},
\end{equation}

\noindent where $N_\parallel(H_2)$ is the column density in units of $10^{21}$
cm$^{-2}$ along a magnetic flux tube and $B_{tot}$ is the total magnetic field
strength in $\mu$G.  For $\mu > 1$, the flux tube is supercritical, meaning it
should collapse due to self-gravity, but for $\mu < 1$ the tube is magnetically
supported.

Because of projection effects between $N_\parallel(H_2)/B_{tot}$ and the
observed quantity $N(H_2)/B_\parallel$, $\mu_{obs}$ will overestimate $\mu$ with
an average correction factor $\mu = \mu_{obs}/3$, assuming a random orientation
of the magnetic field with respect to the line of sight \citep{heiles05}. 
Furthermore, for our data we have measurements of $A_V$, not $N(H_2)$.  The
conversion factor we used before is $N(H_2) = 9.4\times 10^{20} A_V$.  Putting
these two together, Equation \ref{eq:mu1} can be rewritten as:

\begin{equation}
\label{eq:mu2}
\mu = 2.4\: A_V/B_\parallel,
\end{equation}

\noindent with \av{} in magnitudes and \bparallel{} in $\mu$G.

\begin{figure}

\plotone{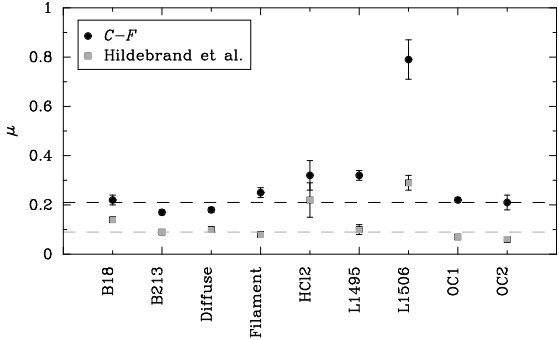}

\caption{\label{fig:mu_crit} Plot of the magnetic critical index, $\mu$, from
Table \ref{tab:bfield} for each subregion.  The weighted average values
(excluding L1506) for the \cf{} and Hildebrand et al.\ techniques are plotted as
black and gray dashed lines, respectively.}

\end{figure}

Using the \av{} values from Tables \ref{tab:poldiffuse} and
\ref{tab:polfilament}, we compute the weighted average \av{} and the weighted
standard deviation of the mean in each subregion. The standard deviation of the
mean is larger than the statistical uncertainty in all subregions.  These
average values are listed in Table \ref{tab:regions}.  Using Equation
\ref{eq:mu2}, we compute $\mu$ for each subregion with both the \cf{} and
Hildebrand et al.\ estimates of \bparallel.  The results are listed in Table
\ref{tab:bfield} and plotted in Figure \ref{fig:mu_crit}.

All subregions are subcritical.  The average $\mu$, excluding L1506, is 0.21
(\cf) or 0.09 (Hildebrand et al).  It is remarkable that $\mu$ is roughly
constant across subregions.  This constancy implies the bulk of the cloud is not
involved with star formation.  In \S\,\ref{sec:align} we argued that our
polarization measurements probed up to at least a column density $A_V \sim 9$
mag.  Since even subregions with known protostars are highly subcritical, it
appears our measured polarizations and estimates of \bparallel{} may not be
probing the innermost regions where stars are forming. However, the connection
between the two is important. The magnetic critical index only regulates
collapse due to ambipolar diffusion.  It is still possible for gas to collapse
\emph{along} the field lines.  The B213 filament appears to be a prime candidate
for this scenario since the magnetic field is parallel to the short axis of the
filament along its entire length.  Small cores within the filament, not probed
by our observations, would then be able to form stars.

\subsection{ Turbulence\label{sec:turbulence}}

In \S\,\ref{sec:align} we argued that our observed power law in the Filament
region, $p \propto A_V^{\phantom{V}0.52\pm0.04}$, was due to the Radiative
Torques model for grain alignment.  \citet{whittet08} did some simple modeling
to show that  RT was effective even at high \av.  However, they did not consider
the possible effects of turbulence, which may also explain this behavior.
\citet[hereafter JKD]{jones92} combined data from various lines-of-sight to
obtain polarization at $K$ ($2.2\:\mu$m) versus optical depth $\tau_K$.  Using
the relation $\tau_K = 0.09 A_V$ from JKD, we fit their data to a power law to
obtain $p_K \propto A_V^{\phantom{V}0.79}$, which is steeper than our observed
power law.  JKD did not model the grain alignment mechanism, but did include the
effects of turbulence with two models.  Their wave model considered a
superposition of cloudlets along the line-of-sight, each threaded by an
Alv{\'e}n wave with random phase and plane of vibration.  All waves had the same
direction of propagation.  Their component model was an extension of the model
of \citet{myers91}.  This model assumed a magnetic field with a constant uniform
component plus a random component that could have arbitrary angles for different
turbulent cells along the line-of-sight.  JKD found that both models best fit
their data when equipartition existed between the magnetic and turbulent energy
densities (wave model) or when the strengths of the uniform and random
components of the magnetic field are equal (component model).  Both of their
models predict that we should measure stronger turbulence because our power law
index is lower than theirs.

We test the predictions of JKD by computing the relevant ratios for the wave and
component models in the Filament region from the values listed in Tables
\ref{tab:regions} and \ref{tab:bfield}.  For the wave model, the turbulent to
magnetic energy density ratio can be written as the ratio of the turbulent to
Alv{\'e}n velocity, $\sigma(v)/V_A$, where $V_A = B/\sqrt{4\pi\rho}$.  Using the
\thirteenco{} velocity dispersion and \bparallel{} from the Hildebrand et al.\
method, we find that $\sigma(v)/V_A = 0.39\pm0.04$.  For the component model,
the ratio of the turbulent-to-large-scale field strength, \bturb, is  exactly
computed by the Hildebrand et al.\ method via Equation \ref{eq:bt}.  In the
Filament, $\langle B_t^2\rangle ^{1/2}/B_0 = 0.20\pm0.02$. Therefore, both
ratios are significantly less than 1 and turbulence is less important than
magnetic fields.  This finding contradicts the predictions of the JKD models. 
Both JKD models assume a constant magnetic field component that does not vary in
angle, and all dispersion in angle is due to turbulence.  However, we know this
is not the  case in Taurus.  The Filament region analyzed has a broad curved
shape that does not appear to be caused by turbulence.  In the next section we
examine possible causes for this morphology.

\begin{figure*}
\epsscale{1}
\plotone{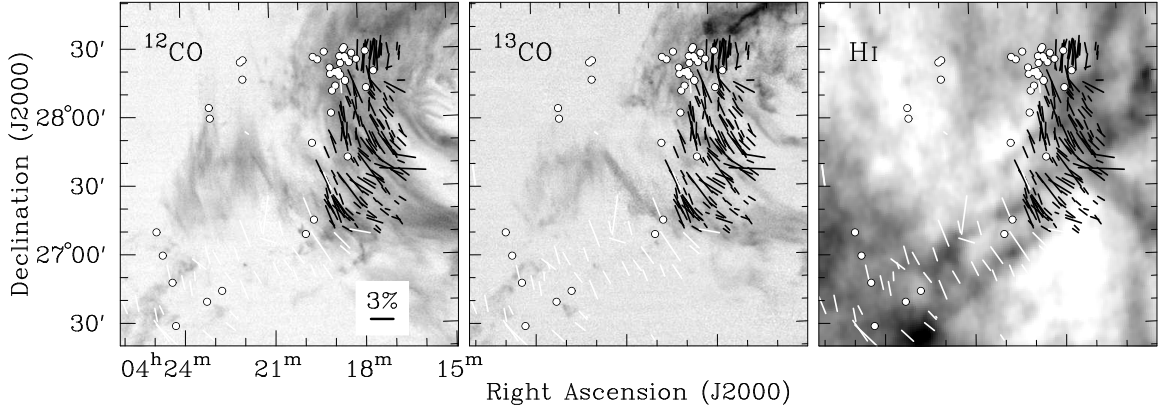}

\caption{\label{fig:map-bubble} Polarization map in the Filament region overlaid
on different background images.  \textit{(Left)} and \textit{(middle)} -
\twelveco{} and \thirteenco{} emission, respectively, from \citet{goldsmith08}. 
\textit{(right)} - \hone{} from M.\ Kr\v{co} (private comm.).  All three
emission images were integrated from $\sim7.6-8.9$ km s$^{-1}$. The vectors and
white circles are the same as in Fig.\ \ref{fig:map-filament}.}

\end{figure*}

\subsection{ Magnetic Field Morphology\label{sec:morphology}}

Figure \ref{fig:map-filament} shows the B213 filament and L1495 region with
polarization vectors overlaid.  The filament is revealed by the \thirteenco{}
emission, which serves as a dense gas tracer.  In B213, the magnetic field is
perpendicular to the apparent long axis of the filament. This morphology can be
readily explained if the gas and dust have gravitationally collapsed along the
field lines to form the filament. Between B213 and L1495, the filament turns
sharply and becomes oriented approximately north-south. The magnetic field also
abruptly transitions from being perpendicular to being parallel to the filament.
Furthermore, the filament itself appears slightly curved, a curvature that is
well-matched by the magnetic field.

\begin{figure*}

\plotone{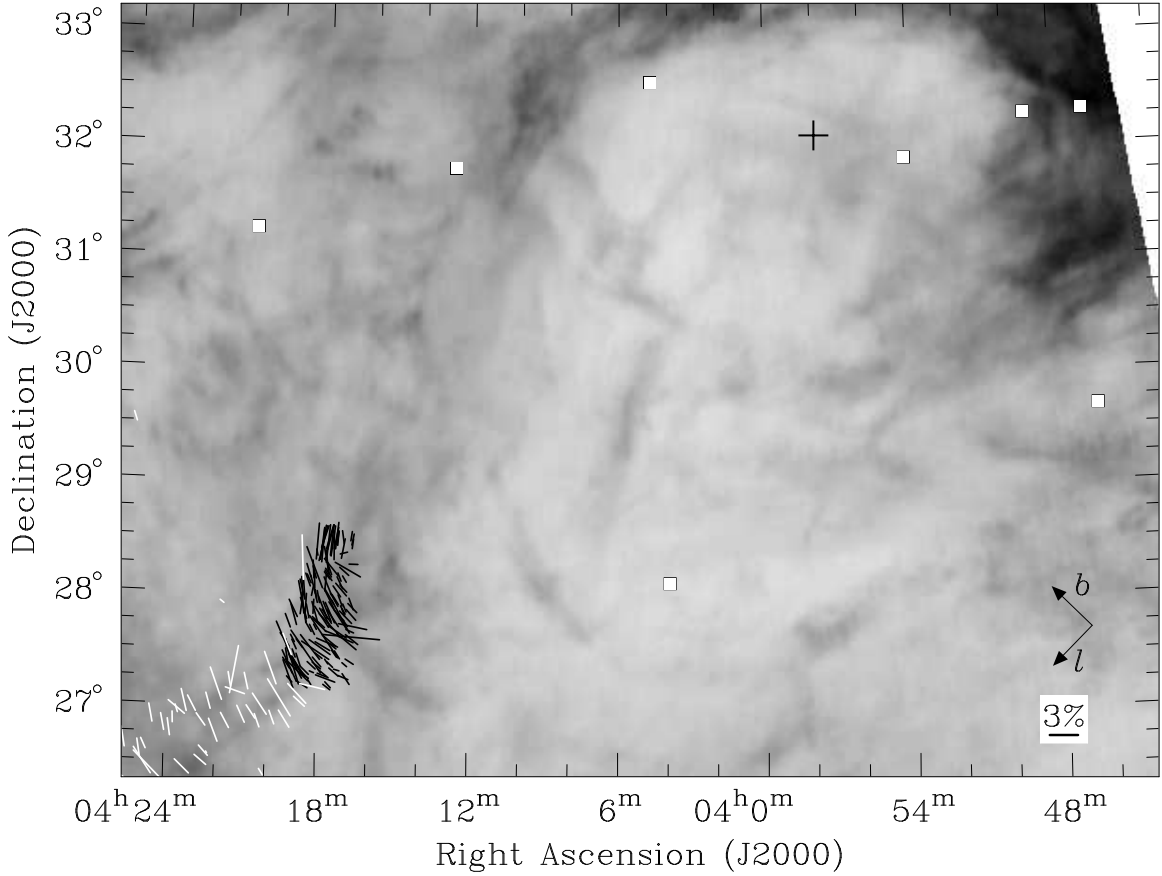}

\caption{\label{fig:map-hi} \hone{} emission integrated from $8.5-9.6$ km
s$^{-1}$ with vectors overlaid as in Fig.\ \ref{fig:map-filament}.  The
squares denote the positions of B stars in the field having parallax
distances between $142-333$ parsecs, while the cross is the position of
gamma-ray pulsar Fermi-LAT PSR J0357+32.}

\end{figure*}

The behavior of the magnetic field in L1495 can be explained if this region is
the compression front of a bubble.  It would explain the curved nature and
the fact that the magnetic field and gas are parallel to each other.  Under
conditions of flux-freezing, an expanding bubble will carry the magnetic field
with it and stretch it to be parallel to the edge of the compression front
\citep[e.g.,][]{novak00,li06}.

In Figure \ref{fig:map-bubble} we show \twelveco, \thirteenco, and \hone{} 
emission in L1495 integrated from $7.6-8.9$ km s$^{-1}$.  This velocity range 
shows the curved bubble-like nature of this region more clearly than the $2-9$ 
km s$^{-1}$ range used in the previous figures.  The \twelveco{} and 
\thirteenco{} are from \citet{goldsmith08} and the \hone{} is Arecibo data from 
M.\ Kr\v{co} (PhD Thesis, Cornell University, in preparation).  In each panel, 
a curved structure in the gas is present that matches the location and radius 
of curvature of the polarization vectors.  Note that the B213 filament is not 
evident in the \twelveco{} and \thirteenco{} emission because it emits at lower 
local standard of rest (LSR) velocities.

Figure \ref{fig:map-hi} shows the shell of a bubble-like structure visible in
the  \hone{} emission to the northwest of L1495, outside the area mapped in
\twelveco{} and \thirteenco.  The \hone{} is integrated from $8.5-9.6$ km
s$^{-1}$ to show both the near and far edges of the shell.  This shell is
approximately $7^\circ$ in  diameter, with one edge at L1495 and the other edge
at the top of the map.  The  \hone{} gas at the top edge of the shell is more
redshifted than the gas in  L1495, implying that whatever source has created the
shell is located at a distance comparable to or greater than that of Taurus.

O and B stars may have sufficient stellar fluxes or winds to clear the 
interstellar medium and create the observed bubble.  Using 
SIMBAD\footnote{http://simbad.u-strasbg.fr/simbad/} we select all O and B 
stars within the area of Figure \ref{fig:map-hi} that also have parallax angles 
of $3-7$ milliarcseconds (parallax distances of $142-333$ pc).  No O stars are 
in the field, and only 8 B stars.  These are shown as squares in Figure 
\ref{fig:map-hi}.  None of the B stars are likely candidates for creating the 
bubble.  Their positions are scattered throughout the map, and they are too 
distant.  Only one B star is closer than 200 pc (at 180 pc) compared with the 
$\sim140$ pc distance of Taurus.  This star is at 3$^\mathrm{h}$46$^\mathrm{m}$ 
$+29^\circ40\arcmin$, but the closest B star in angular separation to L1495 
(4$^\mathrm{h}$03$^\mathrm{m}$ $+28^\circ07\arcmin$) is at 300 pc.

Alternatively, the shell could be a supernova remnant.  A recently discovered 
gamma-ray pulsar, Fermi-LAT PSR J0357+32 has been found in the field by 
\citet{abdo09}.  Using the improved source position from \citet{ray11}, we 
marked the position of this source with a cross in Figure \ref{fig:map-hi}.  
This source has been sought in radio with both Arecibo and GBT, but not 
detected \citep{ray11}.  Current models of pulsars have a gamma-ray beam that 
is broader than the radio beam \citep{watters09}, so Earth may simply not be 
within the radio beam cone.  Because this source has only recently been 
discovered, little is known about it.  The improved positional accuracy of 
\citet{ray11} leads to more accurate estimates of the spin-down frequency 
,$\dot{\nu}$, and a characteristic age $\tau = - \nu/\dot{\nu} = 540$ kyr.  As 
a pulsar that is possibly nearby, with a candidate supernova remnant 
interacting with the Taurus molecular cloud, it should be studied further.

\section{Summary and Conclusions}\label{sec:summary}

We have observed near-infrared $H$-band polarization of background starlight
seen through two regions of the Taurus molecular cloud using the Mimir
instrument, mounted on the 1.8m Perkins telescope located near Flagstaff, AZ and
operated by Lowell Observatory.  After data reduction and selection of
high-quality vectors, we obtained 125 vectors in the diffuse region and 162
vectors in the B213/L1495 region.  The latter region had not been previously 
studied in polarization.  Our results are as follows.

\textit{Grain Alignment} - In the B213/L1495 filament we found an increase in 
the percentage polarization, $p$, versus column density, \av.  These data are
well-fit by a power law with $p  \propto A_V^{\phantom{V}0.52\pm0.04}$.  The
Radiative Torques model for dust  grain alignment is able to explain the
observed trend.  Our near-infrared data  probe the magnetic field geometry up to
at least $A_V \sim 9$ mag., not just in a thin surface layer.

\textit{Magnetic Field Strength} - We divided our data, along with the $\sim400$
previously published optical to near-infrared polarization vectors, into nine
subregions within Taurus.  In each subregion we estimated the strength of the
plane-of-the-sky component of the magnetic field, \bparallel, from the
dispersion in polarization angles.  The Chandrasekhar-Fermi technique (\cf) uses
these dispersions directly to estimate \bparallel{} while the Hildebrand et al.\
method accounts for large-scale spatial variations in polarization angle not
arising from turbulence.  In all subregions we find the Hildebrand et al.\
method produces larger estimates for \bparallel{} than the \cf{} method.  The
magnetic field in L1506 appears to have very different angles in the eastern and
western halves leading to very small values for \bparallel.  The Hildebrand et
al.\ method accounts for this variation better than \cf.  Even so, the small
number of vectors in L1506 leads to an estimate for \bparallel{} that appears
anomalously small ($12\:\mu$G) compared to other subregions.

\textit{Cloud Stability} - We used the derived estimates for \bparallel{} to
compute the critical mass to magnetic flux ratio index, $\mu$.  In all
subregions this index is $< 1$, indicating that the clouds are supported by the
magnetic field. Furthermore, $\mu$ is nearly uniform across all the subregions
with an average value of 0.21 (\cf{} method) or 0.09 (Hildebrand et al.\
method).  The constancy of $\mu$ even in high column density regions like L1495
further supports the idea that our near-infrared data may not be probing the
magnetic field in the densest star-forming cores.  However, $\mu$ only governs
gravitational collapse across the field lines.  The B213 filament is a prime
example where it appears the gas and dust have collapsed along the field lines
to form a filament.

\textit{Cloud Turbulence} - The power law fit to our data in B213/L1495 is
flatter than the observed trend seen by \citet{jones92}.  Based on their models
for turbulence, we expected to find a turbulent to magnetic field energy density
ratio, $\sigma(v)/V_A$, greater than 1, and also the ratio of the turbulent to
large-scale magnetic field strength $> 1$.  We calculated both ratios to be $<
1$ for Taurus, implying that turbulence is less important than the  magnetic
field.  The component model of \citet{jones92} does not allow the non-turbulent
component of the magnetic field to vary in angle.  Furthermore, their wave model
does not allow the Alv{\'e}n wave to vary in direction of propagation.  The
Hildebrand et al.\ method allows for non-turbulent variations in angle, and such
behavior is observed in Taurus (see e.g.\ Figure \ref{fig:map-areas}).
Non-turbulent variations in angle of the magnetic field may explain the
difference between the predictions  of \citet{jones92} and the calculated
quantities.

\textit{Magnetic Field Morphology} - In the L1495 region the Mimir data show
that the magnetic field appears to have a large-scale curvature.  If the Mimir
data are tracing the shell of a bubble, its curvature would explain the observed
morphology.  Arc-shaped segments of a shell are seen in \hone{}. We looked for
potential causes of this structure but found no O stars exist in the region at
the appropriate distance and only 8 B stars, none of which seems a likely
candidate based on location and parallax distance.  However, the shell may be a
supernova remnant.  A recently discovered gamma-ray pulsar, Fermi-LAT PSR
J0357+32, is in the field.  It has been searched for, but remains undetected at
radio frequencies.  However, this non-detection may be a consequence of the
radio beam being narrower than the gamma-ray beam.

The Taurus molecular cloud remains an excellent target for understanding the
importance of magnetic fields in molecular clouds.  Near-infrared polarimeters
such as Mimir can map the large-scale magnetic field to moderate optical depth.
In conjunction with upcoming submillimeter polarimeters like ALMA and
SCUBAPOL2,  we will be able to probe the magnetic field direction from the
largest to smallest spatial scales.

\acknowledgements

Part of the research described in this paper was carried out at the Jet
Propulsion Laboratory, California Institute of Technology, under a contract with
the National Aeronautics and Space Administration.  NLC acknowledges support
from NSF grant AST-0909030 awarded to Northwestern University.  DPC acknowledges
support under NSF AST 06-075500 and 09-07790.

Michael Pavel wrote the astrometry software for Mimir data analysis.
April Pinnick developed the instrumental calibration characterizations,
especially the instrumental polarization across the Mimir field of view.

This research has made use of the SIMBAD database, operated at CDS, Strasbourg,
France.  This research was conducted in part using the Mimir instrument,
jointly developed at Boston University and Lowell Observatory and supported by
NASA, NSF, and the W.M.\ Keck Foundation.

Perkins telescope time for this project was awarded under the Boston
University-Lowell Observatory partnership.  Brian Taylor played key roles in the
smooth operations of both Mimir and the Perkins telescope.

\clearpage

\LongTables
\input{tabpoldiffuse.tex}
\clearpage
\input{tabpolfilament.tex}
\clearpage

\end{document}

%% file: tabregions.tex
\begin{deluxetable}{lcccc}
\tablewidth{0pt}
\tablenum{3}
\tablecaption{Average Physical Parameters for Subregions }\label{tab:regions}

\tablehead{\colhead{} & \colhead{} & \colhead{} & \multicolumn{2}{c}{Velocity Dispersion}\\
\cline{4-5}
\colhead{} & \colhead{\av} & \colhead{$n(H_2)$} & \colhead{\twelveco} & 
\colhead{\thirteenco}\\
\colhead{Subregion} & \colhead{(mag)} & \colhead{(cm$^{-3}$)} &
\colhead{(km s$^{-1}$)} & \colhead{(km s$^{-1}$)}}

\startdata
B18                       & $1.80\pm0.10$ & $660\pm50$ & $1.20\pm0.02$ & $0.90\pm0.02$ \\
B213                      & $2.05\pm0.14$ & $750\pm30$ & $1.16\pm0.02$ & $0.85\pm0.01$ \\
Diffuse\tablenotemark{a}  & $1.30\pm0.03$ & $200\pm10$ & $0.84\pm0.02$ & $0.61\pm0.02$ \\
Filament\tablenotemark{a} & $2.87\pm0.15$ & $920\pm20$ & $1.30\pm0.03$ & $0.98\pm0.02$ \\
HCl2                      & $5.61\pm0.90$ & $970\pm30$ & $1.14\pm0.02$ & $0.76\pm0.01$ \\
L1495                     & $3.32\pm0.18$ & $990\pm10$ & $1.19\pm0.02$ & $0.81\pm0.02$ \\
L1506\tablenotemark{b}    & $1.50\pm0.14$ & $620\pm60$ & $0.87\pm0.01$ & $0.52\pm0.01$ \\
OC1                       & $1.14\pm0.03$ & $210\pm10$ & $0.93\pm0.02$ & $0.71\pm0.01$ \\
OC2                       & $0.91\pm0.09$ & $360\pm50$ & $1.34\pm0.04$ & $0.83\pm0.03$
\enddata

\tablenotetext{a}{Region only contains Mimir data}
\tablenotetext{b}{\mbox{L1506} has two velocity components separated by $\sim$3.5 km
s$^{-1}$.  The velocity dispersion listed here is the average of the dispersion
from each component.}

\end{deluxetable}

%% file: tabbfield.tex
\begin{deluxetable*}{lr@{ $\pm$ }r c c r@{ $\pm$ }r c ccc}
\tablewidth{0pt}
\tablenum{4}
\tablecaption{Magnetic Parameters for Subregions }\label{tab:bfield} 

\tablehead{\colhead{} & \colhead{} & \colhead{} & \colhead{} & \colhead{} &
\multicolumn{3}{c}{$B_\parallel$} & \colhead{} & \multicolumn{2}{c}{$\mu$}\\
\cline{6-8}
\cline{10-11}
\colhead{} & \multicolumn{2}{c}{$b^2$} & \colhead{\# of} & 
\colhead{$\sigma(\theta)$} &
\multicolumn{2}{c}{\cf} & \colhead{Hildebrand} & \colhead{} & 
\colhead{\cf} & \colhead{Hildebrand} \\
\colhead{Subregion} & \multicolumn{2}{c}{(deg$^2$)} & \colhead{vectors} & 
\colhead{(deg)} &
\multicolumn{2}{c}{($\mu$G)} & \colhead{($\mu$G)} &
\colhead{} & \colhead{} & \colhead{}}

\startdata
B18                       & 527 &  52 & $\phantom{1}62$ & 26 &  19 & 1 & $30\pm\phantom{1}2$ & & $0.22\pm0.02$ & $0.14\pm0.01$\\
B213                      & 192 &  24 & $142$           & 18 &  28 & 1 & $51\pm\phantom{1}4$ & & $0.17\pm0.01$ & $0.09\pm0.01$\\
Diffuse\tablenotemark{a}  &  75 &  18 & $125$           & 11 &  17 & 1 & $31\pm\phantom{1}4$ & & $0.18\pm0.01$ & $0.10\pm0.01$\\
Filament\tablenotemark{a} & 124 &  32 & $162$           & 24 &  27 & 1 & $82\pm11$           & & $0.25\pm0.02$ & $0.08\pm0.01$\\
HCl2                      & 144 &  80 & $\phantom{1}22$ & 13 &  42 & 4 & $61\pm17$           & & $0.32\pm0.06$ & $0.22\pm0.07$\\
L1495                     & 103 &  32 & $\phantom{1}99$ & 23 &  25 & 1 & $77\pm12$           & & $0.32\pm0.02$ & $0.10\pm0.02$\\
L1506                     & 921 &   4 & $\phantom{1}39$ & 63 &   5 & 1 & $12\pm\phantom{1}1$ & & $0.79\pm0.08$ & $0.29\pm0.03$\\
OC1                       &  73 &  23 & $191$           & 18 &  12 & 1 & $37\pm\phantom{1}6$ & & $0.22\pm0.01$ & $0.07\pm0.01$\\
OC2                       & 158 &  38 & $\phantom{1}40$ & 34 &  10 & 1 & $38\pm\phantom{1}6$ & & $0.21\pm0.03$ & $0.06\pm0.01$
\enddata

\tablenotetext{a}{Region only contains Mimir data}

\end{deluxetable*}

%% file: tabpoldiffuse.tex
\begin{deluxetable}{rrr r@{ $\pm$ }r r@{ $\pm$ }r r@{ $\pm$ }r}
\tablewidth{0pt}
\tablenum{1}
\tablecaption{$H$-band Polarization Data for Stars in 
the Diffuse Region}\label{tab:poldiffuse}

\tablehead{\colhead{Star} & \colhead{$\alpha$} & \colhead{$\delta$} & 
\multicolumn{2}{c}{$p$} & \multicolumn{2}{c}{$\theta$\tablenotemark{a}} &
\multicolumn{2}{c}{$A_V$\tablenotemark{b}}\\
\colhead{\#} & \colhead{(J2000)} & \colhead{(J2000)} & 
\multicolumn{2}{c}{(\%)} & \multicolumn{2}{c}{(Deg.)} & 
\multicolumn{2}{c}{(mag)}}

\startdata

  1 & 04 50 58.1 & 26 33 16 & 3.18 & 0.60 & $ 50$ &  5 &  1.66 & 0.26\\ 
  2 & 04 50 59.0 & 26 18 15 & 1.82 & 0.15 & $ 41$ &  2 &  1.32 & 0.23\\ 
  3 & 04 50 59.4 & 26 19 40 & 1.57 & 0.23 & $ 41$ &  4 &  1.23 & 0.22\\ 
  4 & 04 51 00.8 & 26 27 40 & 1.16 & 0.18 & $ 49$ &  4 &  1.49 & 0.24\\ 
  5 & 04 51 01.1 & 26 58 30 & 2.25 & 0.58 & $ 43$ &  7 &  1.36 & 0.23\\ 
  6 & 04 51 01.7 & 26 23 54 & 1.67 & 0.27 & $ 51$ &  5 &  1.37 & 0.20\\ 
  7 & 04 51 01.7 & 25 55 22 & 1.64 & 0.17 & $ 39$ &  3 &  1.71 & 0.23\\ 
  8 & 04 51 02.1 & 26 05 47 & 1.56 & 0.33 & $ 44$ &  6 &  1.32 & 0.25\\ 
  9 & 04 51 03.8 & 25 25 39 & 1.18 & 0.27 & $ 31$ &  7 &  1.88 & 0.28\\ 
 10 & 04 51 04.1 & 26 43 46 & 2.21 & 0.54 & $ 46$ &  7 &  1.63 & 0.22\\ 
 11 & 04 51 04.7 & 25 38 31 & 1.88 & 0.16 & $ 36$ &  2 &  1.45 & 0.21\\ 
 12 & 04 51 05.1 & 26 20 07 & 1.60 & 0.25 & $ 49$ &  4 &  1.28 & 0.20\\ 
 13 & 04 51 05.4 & 26 28 28 & 2.26 & 0.28 & $ 47$ &  4 &  1.49 & 0.24\\ 
 14 & 04 51 06.0 & 26 29 27 & 2.77 & 0.83 & $ 43$ &  9 &  1.55 & 0.26\\ 
 15 & 04 51 06.0 & 26 32 00 & 2.47 & 0.10 & $ 51$ &  1 &  1.29 & 0.26\\ 
 16 & 04 51 07.1 & 26 13 53 & 2.23 & 0.67 & $ 56$ &  9 &  1.30 & 0.30\\ 
 17 & 04 51 07.4 & 26 12 54 & 1.76 & 0.12 & $ 38$ &  2 &  1.28 & 0.31\\ 
 18 & 04 51 07.6 & 25 40 22 & 1.07 & 0.16 & $ 36$ &  4 &  1.60 & 0.22\\ 
 19 & 04 51 09.4 & 26 33 06 & 3.08 & 0.57 & $ 45$ &  5 &  1.07 & 0.27\\ 
 20 & 04 51 10.2 & 26 28 37 & 2.08 & 0.68 & $ 34$ &  9 &  1.34 & 0.22\\ 
 21 & 04 51 10.5 & 25 57 17 & 1.88 & 0.42 & $ 55$ &  6 &  1.58 & 0.22\\ 
 22 & 04 51 10.7 & 26 23 44 & 2.13 & 0.51 & $ 52$ &  7 &  1.12 & 0.23\\ 
 23 & 04 51 10.9 & 25 37 21 & 1.10 & 0.09 & $ 35$ &  2 &  1.33 & 0.22\\ 
 24 & 04 51 10.9 & 27 02 26 & 2.42 & 0.28 & $ 59$ &  3 &  1.59 & 0.24\\ 
 25 & 04 51 11.4 & 26 49 10 & 1.71 & 0.48 & $ 44$ &  8 &  1.12 & 0.22\\ 
 26 & 04 51 14.4 & 25 30 23 & 0.59 & 0.16 & $ 42$ &  8 &  1.43 & 0.25\\ 
 27 & 04 51 16.8 & 26 38 51 & 2.14 & 0.29 & $ 49$ &  4 &  1.43 & 0.23\\ 
 28 & 04 51 19.0 & 26 55 36 & 3.84 & 1.20 & $ 60$ &  9 &  1.44 & 0.28\\ 
 29 & 04 51 20.6 & 26 03 36 & 1.44 & 0.21 & $ 42$ &  4 &  1.70 & 0.24\\ 
 30 & 04 51 20.6 & 26 12 06 & 1.28 & 0.19 & $ 41$ &  4 &  0.93 & 0.26\\ 
 31 & 04 51 22.9 & 26 37 07 & 3.60 & 0.99 & $ 54$ &  8 &  1.30 & 0.22\\ 
 32 & 04 51 23.1 & 26 35 28 & 2.42 & 0.06 & $ 51$ &  1 &  1.21 & 0.22\\ 
 33 & 04 51 23.6 & 26 41 25 & 2.53 & 0.17 & $ 50$ &  2 &  1.04 & 0.22\\ 
 34 & 04 51 25.1 & 26 16 07 & 2.34 & 0.76 & $ 57$ &  9 &  0.34 & 0.28\\ 
 35 & 04 51 27.1 & 25 51 20 & 1.82 & 0.21 & $ 47$ &  3 &  1.41 & 0.24\\ 
 36 & 04 51 27.4 & 26 35 57 & 2.55 & 0.62 & $ 51$ &  7 &  1.03 & 0.23\\ 
 37 & 04 51 27.4 & 25 54 39 & 1.47 & 0.10 & $ 44$ &  2 &  1.28 & 0.22\\ 
 38 & 04 51 27.8 & 25 48 55 & 0.84 & 0.13 & $ 37$ &  4 &  1.48 & 0.24\\ 
 39 & 04 51 29.1 & 27 00 15 & 2.02 & 0.36 & $ 59$ &  5 &  2.12 & 0.28\\ 
 40 & 04 51 29.3 & 25 51 43 & 2.00 & 0.37 & $ 37$ &  5 &  1.41 & 0.24\\ 
 41 & 04 51 29.7 & 26 22 23 & 1.05 & 0.27 & $ 52$ &  7 &  1.15 & 0.25\\ 
 42 & 04 51 29.8 & 25 42 47 & 1.49 & 0.49 & $ 28$ &  9 &  1.74 & 0.21\\ 
 43 & 04 51 30.8 & 26 44 01 & 2.38 & 0.45 & $ 52$ &  5 &  1.07 & 0.23\\ 
 44 & 04 51 30.8 & 26 36 37 & 3.20 & 0.73 & $ 44$ &  7 &  1.03 & 0.23\\ 
 45 & 04 51 31.0 & 27 03 45 & 2.80 & 0.47 & $ 58$ &  5 &  1.93 & 0.23\\ 
 46 & 04 51 31.4 & 25 42 11 & 1.58 & 0.07 & $ 33$ &  1 &  1.74 & 0.21\\ 
 47 & 04 51 31.5 & 26 46 20 & 3.02 & 0.55 & $ 40$ &  5 &  1.04 & 0.23\\ 
 48 & 04 51 31.8 & 26 47 50 & 2.36 & 0.57 & $ 44$ &  7 &  1.25 & 0.23\\ 
 49 & 04 51 33.3 & 26 57 12 & 2.05 & 0.11 & $ 53$ &  2 &  1.95 & 0.27\\ 
 50 & 04 51 34.1 & 26 56 18 & 2.06 & 0.29 & $ 48$ &  4 &  1.69 & 0.28\\ 
 51 & 04 51 34.2 & 26 52 35 & 1.52 & 0.34 & $ 55$ &  6 &  1.16 & 0.26\\ 
 52 & 04 51 34.9 & 26 23 05 & 1.71 & 0.52 & $ 59$ &  9 &  1.15 & 0.25\\ 
 53 & 04 51 36.7 & 25 57 54 & 2.32 & 0.63 & $ 56$ &  8 &  1.25 & 0.24\\ 
 54 & 04 51 36.8 & 26 00 54 & 2.21 & 0.68 & $ 47$ &  9 &  1.51 & 0.26\\ 
 55 & 04 51 37.1 & 25 58 52 & 0.94 & 0.05 & $ 37$ &  2 &  1.33 & 0.25\\ 
 56 & 04 51 38.9 & 26 51 24 & 1.73 & 0.18 & $ 52$ &  3 &  1.13 & 0.21\\ 
 57 & 04 51 40.3 & 26 43 43 & 1.76 & 0.35 & $ 52$ &  6 &  1.12 & 0.23\\ 
 58 & 04 51 40.5 & 26 46 17 & 2.50 & 0.22 & $ 51$ &  2 &  1.11 & 0.24\\ 
 59 & 04 51 41.7 & 25 39 30 & 1.73 & 0.42 & $ 28$ &  7 &  1.51 & 0.25\\ 
 60 & 04 51 41.7 & 26 59 29 & 2.75 & 0.26 & $ 60$ &  3 &  1.93 & 0.26\\ 
 61 & 04 51 41.7 & 26 17 30 & 1.32 & 0.12 & $ 47$ &  3 &  1.29 & 0.20\\ 
 62 & 04 51 41.9 & 26 42 24 & 2.16 & 0.20 & $ 53$ &  3 &  1.05 & 0.23\\ 
 63 & 04 51 42.8 & 25 47 09 & 2.40 & 0.33 & $ 22$ &  4 &  1.76 & 0.26\\ 
 64 & 04 51 42.9 & 25 46 40 & 1.65 & 0.24 & $ 24$ &  4 &  1.62 & 0.26\\ 
 65 & 04 51 42.9 & 26 33 04 & 1.94 & 0.51 & $ 54$ &  8 &  0.46 & 0.26\\ 
 66 & 04 51 43.6 & 26 42 18 & 1.89 & 0.19 & $ 56$ &  3 &  1.05 & 0.23\\ 
 67 & 04 51 45.1 & 26 51 44 & 0.28 & 0.08 & $  3$ &  8 &  0.78 & 0.24\\ 
 68 & 04 51 45.9 & 25 42 14 & 1.29 & 0.20 & $ 33$ &  5 &  1.76 & 0.24\\ 
 69 & 04 51 46.5 & 26 27 40 & 2.24 & 0.64 & $ 52$ &  8 &  0.59 & 0.21\\ 
 70 & 04 51 47.1 & 26 51 11 & 1.97 & 0.18 & $ 49$ &  3 &  0.94 & 0.22\\ 
 71 & 04 51 48.7 & 26 12 45 & 1.66 & 0.10 & $ 50$ &  2 &  1.36 & 0.26\\ 
 72 & 04 51 49.3 & 26 20 33 & 1.60 & 0.11 & $ 44$ &  2 &  1.25 & 0.22\\ 
 73 & 04 51 51.4 & 26 12 03 & 1.48 & 0.28 & $ 54$ &  6 &  1.31 & 0.24\\ 
 74 & 04 51 51.8 & 25 35 47 & 2.08 & 0.51 & $ 32$ &  7 &  1.31 & 0.21\\ 
 75 & 04 51 52.7 & 25 32 54 & 0.74 & 0.17 & $ 66$ &  7 &  1.11 & 0.20\\ 
 76 & 04 51 53.0 & 27 03 17 & 2.67 & 0.58 & $ 55$ &  6 &  1.62 & 0.21\\ 
 77 & 04 51 53.1 & 27 02 12 & 1.95 & 0.23 & $ 58$ &  3 &  1.78 & 0.22\\ 
 78 & 04 51 53.3 & 27 02 47 & 2.13 & 0.45 & $ 56$ &  6 &  1.78 & 0.22\\ 
 79 & 04 51 53.7 & 26 57 17 & 2.25 & 0.15 & $ 57$ &  2 &  1.45 & 0.24\\ 
 80 & 04 51 54.2 & 26 49 52 & 2.19 & 0.53 & $ 49$ &  7 &  1.15 & 0.23\\ 
 81 & 04 51 54.3 & 26 41 25 & 2.45 & 0.52 & $ 52$ &  6 &  1.16 & 0.21\\ 
 82 & 04 51 54.4 & 26 25 27 & 1.52 & 0.22 & $ 54$ &  4 &  0.73 & 0.23\\ 
 83 & 04 51 54.9 & 26 47 28 & 2.81 & 0.59 & $ 54$ &  6 &  1.03 & 0.22\\ 
 84 & 04 51 55.9 & 25 46 26 & 1.79 & 0.47 & $ 18$ &  7 &  1.65 & 0.25\\ 
 85 & 04 51 56.8 & 26 06 20 & 1.14 & 0.10 & $ 40$ &  2 &  0.91 & 0.21\\ 
 86 & 04 51 57.2 & 26 49 10 & 2.81 & 0.87 & $ 38$ &  9 &  0.90 & 0.23\\ 
 87 & 04 51 57.3 & 26 39 44 & 2.40 & 0.17 & $ 55$ &  2 &  1.29 & 0.22\\ 
 88 & 04 51 57.7 & 26 40 28 & 2.85 & 0.46 & $ 60$ &  5 &  1.26 & 0.23\\ 
 89 & 04 51 58.2 & 27 02 29 & 1.56 & 0.31 & $ 59$ &  6 &  1.65 & 0.20\\ 
 90 & 04 51 59.4 & 25 46 08 & 2.19 & 0.35 & $ 22$ &  5 &  1.65 & 0.25\\ 
 91 & 04 51 59.9 & 26 41 22 & 2.40 & 0.35 & $ 52$ &  4 &  1.26 & 0.23\\ 
 92 & 04 51 59.9 & 25 59 06 & 2.20 & 0.48 & $ 38$ &  6 &  1.48 & 0.24\\ 
 93 & 04 52 00.5 & 26 38 36 & 2.04 & 0.41 & $ 70$ &  6 &  1.29 & 0.22\\ 
 94 & 04 52 01.3 & 26 48 59 & 1.81 & 0.39 & $ 44$ &  6 &  0.90 & 0.23\\ 
 95 & 04 52 02.3 & 26 57 28 & 2.01 & 0.19 & $ 53$ &  3 &  1.34 & 0.24\\ 
 96 & 04 52 02.3 & 26 54 53 & 2.12 & 0.68 & $ 45$ &  9 &  1.04 & 0.21\\ 
 97 & 04 52 04.2 & 25 44 11 & 1.29 & 0.20 & $ 22$ &  4 &  1.73 & 0.22\\ 
 98 & 04 52 05.4 & 26 01 54 & 1.72 & 0.27 & $ 47$ &  4 &  1.56 & 0.20\\ 
 99 & 04 52 06.1 & 25 32 00 & 0.84 & 0.25 & $ 58$ &  9 &  1.14 & 0.22\\ 
100 & 04 52 06.7 & 25 37 12 & 0.75 & 0.21 & $ 58$ &  8 &  0.97 & 0.25\\ 
101 & 04 52 06.8 & 26 24 56 & 1.39 & 0.43 & $ 48$ &  9 &  0.78 & 0.23\\ 
102 & 04 52 07.1 & 26 18 42 & 1.28 & 0.17 & $ 54$ &  4 &  1.32 & 0.22\\ 
103 & 04 52 07.1 & 25 33 48 & 0.65 & 0.14 & $ 47$ &  6 &  1.16 & 0.22\\ 
104 & 04 52 07.7 & 26 02 30 & 1.94 & 0.22 & $ 38$ &  3 &  1.56 & 0.20\\ 
105 & 04 52 08.3 & 26 13 52 & 1.49 & 0.23 & $ 46$ &  4 &  1.34 & 0.25\\ 
106 & 04 52 08.7 & 25 43 36 & 1.20 & 0.07 & $ 34$ &  2 &  1.63 & 0.24\\ 
107 & 04 52 09.5 & 25 39 50 & 0.78 & 0.09 & $ 40$ &  3 &  1.17 & 0.25\\ 
108 & 04 52 09.8 & 26 23 03 & 1.36 & 0.12 & $ 55$ &  3 &  1.10 & 0.22\\ 
109 & 04 52 11.2 & 27 02 15 & 2.22 & 0.32 & $ 56$ &  4 &  1.56 & 0.22\\ 
110 & 04 52 12.1 & 26 03 39 & 1.34 & 0.11 & $ 41$ &  2 &  1.59 & 0.21\\ 
111 & 04 52 12.3 & 26 52 06 & 2.20 & 0.67 & $ 61$ &  9 &  0.90 & 0.22\\ 
112 & 04 52 12.8 & 26 07 35 & 1.03 & 0.25 & $ 40$ &  7 &  1.48 & 0.24\\ 
113 & 04 52 12.9 & 26 35 53 & 1.92 & 0.20 & $ 58$ &  3 &  1.05 & 0.23\\ 
114 & 04 52 13.1 & 26 25 02 & 1.76 & 0.51 & $ 53$ &  8 &  0.72 & 0.22\\ 
115 & 04 52 13.6 & 26 31 21 & 1.76 & 0.17 & $ 61$ &  3 &  0.92 & 0.23\\ 
116 & 04 52 13.9 & 27 00 38 & 2.28 & 0.44 & $ 52$ &  6 &  1.41 & 0.26\\ 
117 & 04 52 14.6 & 25 55 24 & 0.99 & 0.27 & $ 43$ &  8 &  2.21 & 0.25\\ 
118 & 04 52 15.8 & 26 20 53 & 1.13 & 0.15 & $ 68$ &  4 &  1.23 & 0.20\\ 
119 & 04 52 16.3 & 26 22 03 & 1.40 & 0.24 & $ 66$ &  5 &  1.05 & 0.20\\ 
120 & 04 52 16.6 & 26 54 04 & 2.26 & 0.25 & $ 50$ &  3 &  0.97 & 0.23\\ 
121 & 04 52 17.1 & 26 39 18 & 1.39 & 0.40 & $ 62$ &  8 &  0.82 & 0.26\\ 
122 & 04 52 17.1 & 26 44 01 & 1.55 & 0.47 & $ 58$ &  9 &  0.79 & 0.24\\ 
123 & 04 52 17.8 & 26 51 26 & 2.02 & 0.44 & $ 53$ &  6 &  0.90 & 0.22\\ 
124 & 04 52 20.7 & 25 53 03 & 2.81 & 0.78 & $ 29$ &  8 &  2.57 & 0.28\\ 
125 & 04 52 22.8 & 26 55 59 & 1.47 & 0.41 & $ 62$ &  8 &  0.96 & 0.23

\enddata
\tablecomments{Right Ascension, $\alpha$, is given as hours, minutes, and 
seconds and Declination, $\delta$, is given as degrees, minutes, seconds.}

\tablenotetext{a}{Angles are equatorial, measured east from north.}
\tablenotetext{b}{Extinction map from \citet{pineda10}}
\end{deluxetable}

%% file: tabpolfilament.tex
\begin{deluxetable}{rrr r@{ $\pm$ }r r@{ $\pm$ }r r@{ $\pm$ }r}
\tablewidth{0pt}
\tablenum{2}
\tablecaption{$H$-band Polarization Data for Stars in
the Filament Region}\label{tab:polfilament}

\tablehead{\colhead{Star} & \colhead{$\alpha$} & \colhead{$\delta$} & 
\multicolumn{2}{c}{$p$} & \multicolumn{2}{c}{$\theta$\tablenotemark{a}} &
\multicolumn{2}{c}{$A_V$\tablenotemark{b}}\\
\colhead{\#} & \colhead{(J2000)} & \colhead{(J2000)} & 
\multicolumn{2}{c}{(\%)} & \multicolumn{2}{c}{(Deg.)} & 
\multicolumn{2}{c}{(mag)}}

\startdata

  1 & 04 16 36.5 & 27 42 59 & 3.06 & 0.89 & $ 79$ &  8 &  1.17 & 0.37\\ 
  2 & 04 16 37.9 & 28 16 06 & 0.92 & 0.27 & $ 89$ &  8 &  4.62 & 0.35\\ 
  3 & 04 16 38.1 & 27 52 52 & 1.34 & 0.25 & $ 45$ &  5 &  2.26 & 0.36\\ 
  4 & 04 16 38.3 & 27 36 09 & 1.69 & 0.47 & $ 53$ &  8 &  1.38 & 0.25\\ 
  5 & 04 16 40.2 & 27 37 26 & 6.12 & 1.26 & $ 86$ &  6 &  0.90 & 0.27\\ 
  6 & 04 16 40.4 & 28 29 18 & 1.19 & 0.22 & $-11$ &  5 &  2.59 & 0.33\\ 
  7 & 04 16 40.4 & 28 26 59 & 1.03 & 0.20 & $ -7$ &  6 &  3.06 & 0.27\\ 
  8 & 04 16 41.0 & 27 49 13 & 1.52 & 0.43 & $ 43$ &  8 &  1.96 & 0.35\\ 
  9 & 04 16 41.4 & 27 55 00 & 1.31 & 0.32 & $ 22$ &  7 &  2.68 & 0.41\\ 
 10 & 04 16 42.7 & 27 24 21 & 1.02 & 0.11 & $ 58$ &  3 &  0.30 & 0.33\\ 
 11 & 04 16 43.2 & 27 40 31 & 0.90 & 0.11 & $ 67$ &  4 &  1.66 & 0.33\\ 
 12 & 04 16 43.9 & 28 31 02 & 0.99 & 0.32 & $ -8$ &  9 &  2.75 & 0.31\\ 
 13 & 04 16 50.2 & 27 15 06 & 1.37 & 0.41 & $ 35$ &  9 &  0.42 & 0.28\\ 
 14 & 04 16 50.4 & 27 34 59 & 2.66 & 0.50 & $ 51$ &  5 &  1.57 & 0.26\\ 
 15 & 04 16 51.3 & 28 13 14 & 3.08 & 0.86 & $ 61$ &  8 &  3.44 & 0.40\\ 
 16 & 04 16 52.0 & 27 36 02 & 1.70 & 0.55 & $ 69$ &  9 &  1.28 & 0.28\\ 
 17 & 04 16 53.8 & 27 47 52 & 1.84 & 0.60 & $ 45$ &  9 &  1.63 & 0.29\\ 
 18 & 04 16 54.2 & 27 44 36 & 0.77 & 0.12 & $ 36$ &  4 &  1.80 & 0.31\\ 
 19 & 04 16 54.3 & 27 22 41 & 0.67 & 0.21 & $ 51$ &  9 &  0.46 & 0.32\\ 
 20 & 04 16 56.8 & 27 15 48 & 1.14 & 0.36 & $ 69$ &  9 &  0.41 & 0.29\\ 
 21 & 04 16 57.2 & 27 52 46 & 0.92 & 0.13 & $ 25$ &  4 &  1.38 & 0.29\\ 
 22 & 04 16 58.8 & 28 08 10 & 0.77 & 0.10 & $ 32$ &  4 &  2.23 & 0.29\\ 
 23 & 04 16 59.7 & 27 50 29 & 0.87 & 0.14 & $ 22$ &  5 &  1.91 & 0.26\\ 
 24 & 04 16 59.8 & 28 12 14 & 1.30 & 0.37 & $ 49$ &  8 &  2.92 & 0.35\\ 
 25 & 04 17 00.9 & 28 03 07 & 1.07 & 0.27 & $ 44$ &  7 &  2.48 & 0.29\\ 
 26 & 04 17 00.9 & 27 51 43 & 1.02 & 0.16 & $ 30$ &  4 &  1.58 & 0.28\\ 
 27 & 04 17 01.8 & 28 21 59 & 0.80 & 0.20 & $-74$ &  7 &  5.92 & 0.40\\ 
 28 & 04 17 03.2 & 28 24 20 & 2.10 & 0.61 & $-10$ &  8 &  5.56 & 0.31\\ 
 29 & 04 17 03.8 & 28 30 26 & 1.38 & 0.44 & $  9$ &  9 &  2.51 & 0.29\\ 
 30 & 04 17 06.0 & 27 48 08 & 3.97 & 0.30 & $ 29$ &  2 &  2.11 & 0.28\\ 
 31 & 04 17 06.2 & 28 02 33 & 1.77 & 0.12 & $ 41$ &  2 &  2.54 & 0.29\\ 
 32 & 04 17 08.1 & 27 56 45 & 1.24 & 0.14 & $ 21$ &  3 &  2.03 & 0.28\\ 
 33 & 04 17 09.7 & 28 06 09 & 1.41 & 0.45 & $ 26$ &  9 &  3.08 & 0.29\\ 
 34 & 04 17 09.8 & 27 48 51 & 1.56 & 0.48 & $ 43$ &  9 &  2.77 & 0.32\\ 
 35 & 04 17 10.8 & 28 09 09 & 0.95 & 0.11 & $ 22$ &  3 &  2.51 & 0.27\\ 
 36 & 04 17 12.3 & 27 27 46 & 0.99 & 0.15 & $ 33$ &  4 &  1.17 & 0.34\\ 
 37 & 04 17 12.3 & 27 48 36 & 3.51 & 0.51 & $ 34$ &  4 &  2.77 & 0.32\\ 
 38 & 04 17 12.8 & 27 39 59 & 2.32 & 0.52 & $ 57$ &  6 &  1.81 & 0.32\\ 
 39 & 04 17 13.3 & 27 19 45 & 0.62 & 0.14 & $ 75$ &  7 &  0.52 & 0.32\\ 
 40 & 04 17 16.1 & 28 29 54 & 1.56 & 0.52 & $ -3$ & 10 &  3.31 & 0.27\\ 
 41 & 04 17 16.4 & 28 30 00 & 1.34 & 0.20 & $ -8$ &  4 &  3.31 & 0.27\\ 
 42 & 04 17 16.5 & 28 01 34 & 1.41 & 0.10 & $ 36$ &  2 &  2.65 & 0.33\\ 
 43 & 04 17 18.7 & 28 31 11 & 2.82 & 0.47 & $ -6$ &  5 &  3.21 & 0.29\\ 
 44 & 04 17 19.4 & 27 28 22 & 0.90 & 0.14 & $ 37$ &  5 &  1.02 & 0.31\\ 
 45 & 04 17 19.7 & 27 18 37 & 1.31 & 0.41 & $ 51$ &  9 &  1.05 & 0.31\\ 
 46 & 04 17 19.9 & 27 57 09 & 1.29 & 0.20 & $ 33$ &  4 &  2.15 & 0.28\\ 
 47 & 04 17 20.4 & 28 29 52 & 2.29 & 0.35 & $-10$ &  4 &  3.31 & 0.27\\ 
 48 & 04 17 22.4 & 27 57 04 & 1.62 & 0.27 & $ 31$ &  5 &  2.83 & 0.30\\ 
 49 & 04 17 23.6 & 27 38 57 & 1.76 & 0.33 & $ 70$ &  5 &  2.67 & 0.37\\ 
 50 & 04 17 25.1 & 27 47 17 & 2.19 & 0.46 & $ 35$ &  6 &  3.70 & 0.41\\ 
 51 & 04 17 25.2 & 27 32 37 & 1.11 & 0.09 & $ 53$ &  2 &  1.57 & 0.29\\ 
 52 & 04 17 25.5 & 28 25 54 & 3.02 & 0.20 & $  1$ &  2 &  5.71 & 0.45\\ 
 53 & 04 17 27.9 & 28 28 15 & 3.23 & 0.20 & $ -7$ &  2 &  4.31 & 0.30\\ 
 54 & 04 17 28.8 & 27 22 33 & 0.84 & 0.16 & $ 47$ &  5 &  1.23 & 0.32\\ 
 55 & 04 17 30.6 & 28 28 46 & 3.25 & 0.44 & $ -5$ &  4 &  4.31 & 0.30\\ 
 56 & 04 17 31.0 & 28 14 19 & 2.35 & 0.52 & $ 40$ &  6 &  5.58 & 0.38\\ 
 57 & 04 17 31.8 & 27 35 11 & 1.86 & 0.40 & $ 52$ &  6 &  3.16 & 0.30\\ 
 58 & 04 17 31.9 & 27 50 13 & 2.38 & 0.08 & $ 26$ &  1 &  4.44 & 0.33\\ 
 59 & 04 17 33.1 & 27 11 49 & 0.86 & 0.14 & $ 44$ &  5 &  0.98 & 0.28\\ 
 60 & 04 17 34.1 & 28 30 09 & 2.70 & 0.19 & $-14$ &  2 &  4.90 & 0.28\\ 
 61 & 04 17 34.5 & 27 58 49 & 1.87 & 0.33 & $ 32$ &  5 &  3.88 & 0.34\\ 
 62 & 04 17 34.6 & 27 44 13 & 2.91 & 0.91 & $ 44$ &  9 &  6.67 & 0.39\\ 
 63 & 04 17 34.8 & 27 57 34 & 2.37 & 0.17 & $ 28$ &  2 &  4.32 & 0.33\\ 
 64 & 04 17 35.8 & 27 10 35 & 0.94 & 0.24 & $ 53$ &  7 &  0.87 & 0.31\\ 
 65 & 04 17 36.0 & 28 30 25 & 2.67 & 0.62 & $-15$ &  7 &  4.90 & 0.28\\ 
 66 & 04 17 36.7 & 27 34 11 & 2.43 & 0.31 & $ 68$ &  4 &  2.59 & 0.29\\ 
 67 & 04 17 37.5 & 28 11 23 & 1.09 & 0.17 & $ 36$ &  4 &  5.92 & 0.35\\ 
 68 & 04 17 37.5 & 28 09 57 & 1.32 & 0.29 & $ 35$ &  6 &  6.32 & 0.36\\ 
 69 & 04 17 37.7 & 28 14 58 & 1.96 & 0.40 & $ 21$ &  6 &  7.22 & 0.40\\ 
 70 & 04 17 37.8 & 27 24 08 & 0.60 & 0.14 & $ 36$ &  7 &  1.14 & 0.28\\ 
 71 & 04 17 38.2 & 28 04 14 & 2.10 & 0.28 & $ 24$ &  4 &  4.92 & 0.47\\ 
 72 & 04 17 38.6 & 27 50 47 & 2.29 & 0.42 & $ 23$ &  5 &  4.24 & 0.30\\ 
 73 & 04 17 39.3 & 27 15 05 & 2.20 & 0.63 & $ 71$ &  8 &  0.95 & 0.25\\ 
 74 & 04 17 39.4 & 27 48 14 & 1.82 & 0.49 & $ 35$ &  8 &  5.17 & 0.30\\ 
 75 & 04 17 39.6 & 28 26 52 & 4.02 & 0.51 & $ -5$ &  4 &  5.80 & 0.34\\ 
 76 & 04 17 40.5 & 28 09 53 & 0.97 & 0.15 & $ 39$ &  5 &  6.32 & 0.36\\ 
 77 & 04 17 40.9 & 28 17 16 & 2.47 & 0.35 & $  7$ &  4 &  7.10 & 0.38\\ 
 78 & 04 17 42.6 & 28 18 07 & 2.88 & 0.94 & $ 12$ &  9 &  6.49 & 0.49\\ 
 79 & 04 17 43.2 & 27 47 40 & 2.21 & 0.10 & $ 32$ &  1 &  5.74 & 0.31\\ 
 80 & 04 17 43.9 & 28 05 57 & 1.72 & 0.23 & $ -4$ &  4 &  5.23 & 0.59\\ 
 81 & 04 17 44.6 & 27 17 44 & 1.44 & 0.24 & $ 51$ &  5 &  0.93 & 0.26\\ 
 82 & 04 17 45.3 & 27 37 18 & 5.40 & 0.93 & $ 35$ &  5 &  8.72 & 0.40\\ 
 83 & 04 17 47.6 & 27 32 16 & 1.81 & 0.52 & $ 60$ &  8 &  5.01 & 0.37\\ 
 84 & 04 17 47.6 & 28 30 30 & 2.43 & 0.48 & $-18$ &  6 &  6.26 & 0.29\\ 
 85 & 04 17 47.7 & 27 16 53 & 1.06 & 0.26 & $ 68$ &  7 &  0.93 & 0.26\\ 
 86 & 04 17 49.3 & 27 52 57 & 2.61 & 0.53 & $ 23$ &  6 &  4.43 & 0.34\\ 
 87 & 04 17 51.1 & 28 29 16 & 2.96 & 0.31 & $ -9$ &  3 &  6.72 & 0.31\\ 
 88 & 04 17 51.6 & 27 47 52 & 2.68 & 0.71 & $ 34$ &  8 &  3.53 & 0.28\\ 
 89 & 04 17 51.7 & 28 26 59 & 3.81 & 0.38 & $-12$ &  3 &  6.83 & 0.35\\ 
 90 & 04 17 51.9 & 28 25 51 & 3.52 & 0.30 & $ -4$ &  2 &  7.82 & 0.41\\ 
 91 & 04 17 53.5 & 28 26 50 & 3.42 & 0.53 & $ -6$ &  4 &  6.83 & 0.35\\ 
 92 & 04 17 54.2 & 27 31 44 & 2.73 & 0.40 & $ 56$ &  4 &  5.01 & 0.37\\ 
 93 & 04 17 55.9 & 27 43 25 & 1.98 & 0.63 & $ 47$ &  9 &  1.01 & 0.33\\ 
 94 & 04 17 56.6 & 27 27 10 & 0.93 & 0.22 & $ 53$ &  7 &  3.13 & 0.31\\ 
 95 & 04 17 58.8 & 27 45 00 & 1.75 & 0.30 & $ 49$ &  5 &  1.22 & 0.28\\ 
 96 & 04 18 00.6 & 27 58 32 & 1.30 & 0.15 & $ 12$ &  3 &  2.41 & 0.30\\ 
 97 & 04 18 00.9 & 27 23 46 & 1.38 & 0.40 & $ 42$ &  8 &  1.84 & 0.36\\ 
 98 & 04 18 01.0 & 27 21 55 & 1.13 & 0.32 & $ 52$ &  8 &  1.79 & 0.30\\ 
 99 & 04 18 04.6 & 27 30 01 & 3.46 & 0.67 & $ 57$ &  6 &  7.80 & 0.45\\ 
100 & 04 18 05.4 & 28 28 01 & 3.82 & 0.20 & $ -3$ &  1 &  8.50 & 0.50\\ 
101 & 04 18 05.7 & 28 22 07 & 0.67 & 0.13 & $ 12$ &  6 & 12.85 & 0.65\\ 
102 & 04 18 06.3 & 27 58 51 & 1.42 & 0.26 & $ 13$ &  5 &  2.60 & 0.28\\ 
103 & 04 18 06.5 & 27 59 52 & 0.90 & 0.28 & $  9$ &  9 &  2.71 & 0.27\\ 
104 & 04 18 06.8 & 27 42 30 & 0.61 & 0.18 & $ 28$ &  8 &  2.13 & 0.29\\ 
105 & 04 18 08.5 & 27 50 41 & 2.13 & 0.21 & $ 17$ &  3 &  1.96 & 0.33\\ 
106 & 04 18 10.5 & 28 06 43 & 1.55 & 0.44 & $  3$ &  8 &  3.23 & 0.42\\ 
107 & 04 18 10.9 & 28 11 36 & 1.34 & 0.16 & $ 16$ &  3 &  3.61 & 0.38\\ 
108 & 04 18 11.1 & 28 14 03 & 4.66 & 0.87 & $ 22$ &  5 &  4.55 & 0.39\\ 
109 & 04 18 11.9 & 27 53 17 & 1.51 & 0.46 & $ -2$ &  9 &  2.17 & 0.39\\ 
110 & 04 18 12.3 & 27 18 30 & 0.48 & 0.07 & $ 68$ &  4 &  2.25 & 0.37\\ 
111 & 04 18 13.4 & 27 21 56 & 1.16 & 0.31 & $ 48$ &  8 &  2.45 & 0.36\\ 
112 & 04 18 13.7 & 27 41 01 & 1.19 & 0.15 & $ 28$ &  4 &  2.55 & 0.31\\ 
113 & 04 18 13.9 & 28 04 56 & 2.05 & 0.25 & $ 16$ &  3 &  3.13 & 0.31\\ 
114 & 04 18 14.0 & 28 13 09 & 1.89 & 0.30 & $  4$ &  5 &  4.55 & 0.39\\ 
115 & 04 18 15.2 & 28 01 22 & 0.94 & 0.20 & $-10$ &  6 &  2.78 & 0.28\\ 
116 & 04 18 16.8 & 27 13 02 & 0.64 & 0.07 & $ 33$ &  3 &  0.93 & 0.30\\ 
117 & 04 18 17.4 & 27 31 37 & 3.32 & 0.08 & $ 39$ &  1 &  6.97 & 0.40\\ 
118 & 04 18 22.5 & 27 42 22 & 1.16 & 0.19 & $ 32$ &  5 &  1.14 & 0.27\\ 
119 & 04 18 23.7 & 28 08 05 & 1.02 & 0.22 & $ 25$ &  6 &  2.24 & 0.29\\ 
120 & 04 18 24.8 & 28 12 27 & 1.47 & 0.16 & $ 19$ &  3 &  2.91 & 0.36\\ 
121 & 04 18 25.3 & 27 50 47 & 1.52 & 0.25 & $ 12$ &  5 &  2.29 & 0.37\\ 
122 & 04 18 26.5 & 28 07 28 & 1.44 & 0.36 & $ 21$ &  7 &  2.24 & 0.29\\ 
123 & 04 18 26.7 & 27 16 02 & 0.78 & 0.17 & $ 57$ &  6 &  1.96 & 0.30\\ 
124 & 04 18 27.3 & 27 29 07 & 5.53 & 1.58 & $ 36$ &  8 &  5.17 & 0.42\\ 
125 & 04 18 28.7 & 27 16 48 & 1.06 & 0.35 & $ 47$ &  9 &  2.79 & 0.28\\ 
126 & 04 18 30.0 & 27 19 15 & 1.27 & 0.17 & $ 46$ &  4 &  3.39 & 0.32\\ 
127 & 04 18 30.1 & 27 54 51 & 2.09 & 0.40 & $ -2$ &  5 &  2.06 & 0.26\\ 
128 & 04 18 31.2 & 27 48 15 & 1.36 & 0.18 & $  9$ &  4 &  1.58 & 0.36\\ 
129 & 04 18 34.8 & 28 01 55 & 2.01 & 0.25 & $  6$ &  4 &  2.25 & 0.31\\ 
130 & 04 18 35.2 & 27 11 44 & 1.38 & 0.44 & $ 58$ &  9 &  2.04 & 0.32\\ 
131 & 04 18 35.3 & 27 52 21 & 2.32 & 0.42 & $ 18$ &  5 &  2.29 & 0.28\\ 
132 & 04 18 37.2 & 27 55 00 & 2.17 & 0.57 & $  6$ &  8 &  1.91 & 0.26\\ 
133 & 04 18 38.6 & 27 55 45 & 4.45 & 0.92 & $ 12$ &  6 &  1.91 & 0.26\\ 
134 & 04 18 39.5 & 28 02 28 & 2.09 & 0.19 & $ 15$ &  3 &  2.27 & 0.28\\ 
135 & 04 18 40.1 & 28 04 28 & 1.79 & 0.50 & $  4$ &  8 &  2.02 & 0.29\\ 
136 & 04 18 41.2 & 27 52 45 & 1.42 & 0.37 & $ -7$ &  7 &  1.92 & 0.28\\ 
137 & 04 18 43.4 & 27 33 33 & 1.14 & 0.30 & $ 38$ &  8 &  1.35 & 0.34\\ 
138 & 04 18 45.2 & 27 24 49 & 1.40 & 0.25 & $ 19$ &  5 &  3.63 & 0.34\\ 
139 & 04 18 46.1 & 27 18 48 & 3.05 & 0.28 & $ 47$ &  3 &  6.93 & 0.30\\ 
140 & 04 18 46.3 & 27 21 42 & 2.85 & 0.34 & $ 35$ &  3 &  7.27 & 0.32\\ 
141 & 04 18 47.3 & 27 21 29 & 2.15 & 0.36 & $ 29$ &  5 &  7.27 & 0.32\\ 
142 & 04 18 50.2 & 27 20 31 & 2.58 & 0.42 & $ 30$ &  5 &  5.49 & 0.31\\ 
143 & 04 18 53.7 & 27 23 55 & 2.01 & 0.38 & $ 32$ &  5 &  2.85 & 0.29\\ 
144 & 04 18 55.1 & 27 20 17 & 3.54 & 0.47 & $ 18$ &  4 &  7.16 & 0.31\\ 
145 & 04 18 55.9 & 27 33 00 & 1.12 & 0.08 & $ 25$ &  2 &  1.26 & 0.29\\ 
146 & 04 18 56.0 & 27 28 13 & 0.93 & 0.07 & $ 21$ &  2 &  1.70 & 0.29\\ 
147 & 04 18 57.4 & 27 23 47 & 1.80 & 0.30 & $ 31$ &  5 &  2.31 & 0.28\\ 
148 & 04 18 58.3 & 27 48 52 & 2.27 & 0.60 & $ 16$ &  8 &  1.95 & 0.27\\ 
149 & 04 19 00.2 & 27 36 09 & 0.87 & 0.26 & $ 29$ &  9 &  0.94 & 0.29\\ 
150 & 04 19 00.2 & 27 52 10 & 2.02 & 0.49 & $ 27$ &  7 &  1.93 & 0.31\\ 
151 & 04 19 05.2 & 27 14 21 & 0.87 & 0.11 & $ 42$ &  4 & 13.04 & 0.50\\ 
152 & 04 19 05.8 & 27 21 26 & 3.66 & 0.88 & $ 26$ &  7 &  2.14 & 0.34\\ 
153 & 04 19 06.8 & 27 21 21 & 1.69 & 0.26 & $ 26$ &  4 &  2.14 & 0.34\\ 
154 & 04 19 08.5 & 27 17 09 & 3.15 & 0.40 & $ 21$ &  4 &  7.36 & 0.45\\ 
155 & 04 19 10.0 & 27 29 21 & 1.17 & 0.10 & $ 20$ &  3 &  1.07 & 0.29\\ 
156 & 04 19 12.1 & 27 32 56 & 1.06 & 0.11 & $ 27$ &  3 &  0.79 & 0.25\\ 
157 & 04 19 12.2 & 27 32 42 & 1.25 & 0.10 & $ 16$ &  2 &  0.79 & 0.25\\ 
158 & 04 19 12.2 & 27 45 31 & 1.52 & 0.28 & $ 22$ &  5 &  1.71 & 0.35\\ 
159 & 04 19 13.2 & 27 27 33 & 1.32 & 0.42 & $ 11$ &  9 &  1.40 & 0.27\\ 
160 & 04 19 13.9 & 27 22 59 & 1.35 & 0.18 & $ 33$ &  4 &  1.90 & 0.29\\ 
161 & 04 19 18.8 & 27 35 24 & 1.68 & 0.45 & $ 36$ &  8 &  1.32 & 0.23\\ 
162 & 04 19 23.1 & 27 36 48 & 3.54 & 0.78 & $ 24$ &  6 &  1.32 & 0.23
\enddata
\tablecomments{Right Ascension, $\alpha$, is given as hours, minutes, and 
seconds and Declination, $\delta$, is given as degrees, minutes, seconds.}

\tablenotetext{a}{Angles are equatorial, measured east from north.}
\tablenotetext{b}{Extinction map from \citet{pineda10}}

\end{deluxetable}